\RequirePackage{fix-cm}
\documentclass[onecollarge]{svjour2}       
\smartqed  
\usepackage{graphicx}
\usepackage{caption}
\usepackage{bm}
\usepackage{amssymb}
\usepackage{amsmath}
\usepackage{amsfonts}
\usepackage{mathtools}
\usepackage{mathrsfs}
\usepackage{nccmath}
\usepackage{color}
\usepackage{amsbsy}
\DeclareMathAlphabet{\mathbfscr}{OMS}{mdugm}{b}{n}

\makeatletter
\newcommand*{\defeq}{\mathrel{\rlap{%
                     \raisebox{0.3ex}{$\m@th\cdot$}}%
                     \raisebox{-0.3ex}{$\m@th\cdot$}}%
                     =}
\newcommand*{\eqdef}{=\mathrel{\rlap{%
                     \raisebox{0.3ex}{$\m@th\cdot$}}%
                     \raisebox{-0.3ex}{$\m@th\cdot$}}%
                     }

\makeatother

\newcommand{\IR}{\mathbb{R}}

\newcommand{\CC}{\mathcal{C}}
\newcommand{\CCBGK}{\CC_{\mathrm{BGK}}}
\newcommand{\IE}{\mathbb{E}}
\newcommand{\MM}{\mathcal{M}}
\newcommand{\IM}{\mathbb{M}}
\newcommand{\II}{\mathbb{I}}
\newcommand{\ff}{\mathcal{F}}
\newcommand{\IF}{\mathbb{F}}
\newcommand{\ID}{\mathbb{D}}
\newcommand{\IX}{\mathbb{X}}
\renewcommand{\IJ}{\mathbb{J}}
\newcommand{\FL}{\ff_{\mathrm{L}}}
\newcommand{\FN}{\ff_{N}}
\newcommand{\IA}{\mathbb{A}}

\newcommand{\HH}{\eta}
\newcommand{\Id}{\mathrm{Id}}
\newcommand{\Domain}{\mathscr{D}}
\newcommand{\EQ}[1]{(\ref{eq:#1})}
\newcommand{\SEC}[1]{\ref{sec:#1}}
\newcommand{\feq}{f_{\mathrm{eq}}}

\DeclareMathOperator*{\argmin}{arg\:min}

\begin{document}

\title{Moment closure approximations of the Boltzmann Equation based on $\varphi$-divergences}


\author{M.R.A. Abdel-Malik         \and
        E.H. van Brummelen 
}


\institute{M. Abdel Malik \at
              Department of Mechanical Engineering, Eindhoven University of Technology \\
              \email{mabdel@tue.nl}           
           \and
           E.H. van Brummelen \at Department of Mechanical Engineering \& Department of Mathematics and Computer Science, Eindhoven University of Technology
}

\date{Received: date / Accepted: date}

\maketitle

\begin{abstract}
This paper is concerned with approximations of the Boltzmann equation based on the method of moments. We propose a generalization of the setting
of the moment-closure problem from relative entropy to $\varphi$\nobreakdash-divergences and a corresponding closure procedure based on minimization 
of $\varphi$-divergences. The proposed description encapsulates as special cases Grad's classical closure based on expansion in Hermite polynomials and Levermore's 
entropy-based closure. We establish that the generalization to divergence-based closures enables the construction of extended thermodynamic theories that avoid 
essential limitations of the standard moment-closure formulations such as inadmissibility of the approximate phase-space distribution, potential loss of hyperbolicity
and singularity of flux functions at local equilibrium. The divergence-based closure leads to a hierarchy of tractable symmetric hyperbolic systems that retain the fundamental structural properties of the Boltzmann equation.
\end{abstract}

\section{Introduction}
\label{sec:intro}
The Boltzmann equation provides a description of the molecular dynamics of fluid flows based on their one-particle phase-space distribution. However,
the Boltzmann equation also encapsulates all conventional macroscopic flow models in the sense that its limit solutions correspond to solutions of
the compressible Euler and Navier--Stokes equations~\cite{Bardos:1991vf,Esposito:1994ca}, the incompressible Euler and Navier--Stokes equations~\cite{Golse:2004oe,Lions:2001sj}, the incompressible Stokes equations~\cite{Lions:2001wb} and
the incompressible Navier--Stokes--Fourier system~\cite{Levermore:2010kx}; see~\cite{Saint-Raymond2009} for an overview. 
The Boltzmann equation is uniquely suited to describe flows in 
the transitional molecular/continuum regime and the corresponding rarefaction effects, by virtue of its inherent characterization of deviations of the velocity distribution from local equilibrium. Applications in which rarefaction effects play a significant role are multitudinous, including gas flow problems 
involving large mean free paths in high-altitude flows and hypobaric applications such as chemical vapor deposition; see \cite{Cercignani2000,Struchtrup2005} and references therein for further examples.
Moreover, the perpetual trend toward miniaturization in science and technology renders accurate descriptions of fluid flows in the transitional molecular/continuum regime of fundamental technological relevance, for instance, in nanoscale applications, micro-channel flows or flow in porous media~\cite{shen2005}.  
The Boltzmann equation also provides a prototype for kinetic models in many other applications that require a description of the collective
behavior of large ensembles of small particles, for instance, in semi-conductors~\cite{jungel2009}, in plasmas and fusion and fission devices~\cite{miyamoto2004}  and
in dispersed-particle flows such as in fluidized-bed reactors~\cite{reeks1991,reeks1992,reeks1993}.

Numerical approximation of the Boltzmann equation poses a formidable challenge, on account of its high dimensional setting: for a problem in~$D$ spatial dimensions, the one-particle phase-space is~$2D$ dimensional. The corresponding computational complexity of conventional discretization methods for (integro-)differential equations, such as finite-element methods with uniform meshes, is prohibitive. Numerical approximations of the Boltzmann equations
have been predominantly based on particle methods, such as the Direct Simulation Monte Carlo (DSMC) method~\cite{Bird:1970,Bird:1994}. Convergence proofs for these methods~\cite{Wagner1992} however convey that their computational complexity depends sensitively on the Knudsen number, and the computational cost becomes prohibitive in the fluid-dynamical limit. Moreover, from an approximation perspective, DSMC can be inefficient, because it is inherent to the underlying Monte-Carlo process that the approximation error decays only as $n^{-\frac{1}{2}}$ as the number of simulation molecules,~$n$, increases; see, for instance, \cite[Thm.~5.14]{Klenke2008}. Efficient
computational modeling of fluid flows in the transitional molecular/continuum regime therefore remains an outstanding challenge.

An alternative approximation technique for the Boltzmann equation, which has been only relatively sparsely investigated, is the method of moments \cite{Grad1949,Levermore1996,Struchtrup2005}. The method of moments represents a general statistical approximation technique which identifies parameters of an approximate distribution
based on its moments~\cite{Matyas1999}. Application of the method of moments to the Boltzmann equation engenders an evolution equation for the moments (weighted averages) of the phase-space distribution. An approximation based on moments is generally consistent with a restricted interest in functionals of the distribution 
corresponding to macroscopic properties of the fluid. The method of moments is closely related to extended thermodynamics; see, for instance,~\cite{Dreyer1987,Muller1993}.

Intrinsic to the method of moments is an approximation of the moment-closure relation, viz., a
relation that closes the evolution equation for the moments. Moment-closure approximations for the Boltzmann equation were originally conceived by Grad~\cite{Grad1949}. 
Grad's moment closure is based on an expansion of the one-particle distribution in Hermite polynomials.
For a linear Boltzmann equation extended with an exogenous forcing, Schmeiser and Zwirchmayr~\cite{Schmeiser1998} have shown that the distribution in 
Grad's moment equations converges to the distribution of the underlying kinetic model as the order of the moment approximation tends to infinity, and to the solution of a corresponding drift-diffusion model in the macroscopic limit, i.e. as the Knudsen number tends to zero. Grad's moment systems are impaired by two
essential deficiencies, however, viz., the potential occurrence of inadmissible locally negative phase-space distributions and potential loss of hyperbolicity~\cite{Brini2001,Torrilhon2000}. Levermore~\cite{Levermore1996} has developed a moment-closure procedure based on constrained entropy minimization, similar to Dreyer's maximum-entropy closure in extended thermodynamics~\cite{Dreyer1987}. The entropy minimization procedure formally leads to an exponential closure. Levermore's moment systems retain the fundamental structural properties of the Boltzmann equation, viz., conservation of mass, momentum and energy, Galilean invariance and entropy dissipation. Moreover, the moment systems form a hierarchy of symmetric hyperbolic systems and the corresponding distributions are non-negative. It was later shown by Junk~\cite{Junk1998}, however, that Levermore's moment-closure procedure is impaired by a realizability problem, in that there exist moments for which the minimum-entropy distribution is non-existent. 
Moreover, the fluxes in Levermore's moment systems may become arbitrarily large in the vicinity of (local) equilibrium. 
Recent results by Junk~\cite{Junk2000}, 
Schneider~\cite{Schneider2004} and Pavan~\cite{Pavan2011} convey that potential non-existence of solutions to the entropy-minimization problem can be avoided by relaxing the constraints to allow inequalities in the highest-order moments. Moreover, the solution to the relaxed entropy minimization problem coincides with the solution to the original constrained entropy minimization problem if the latter admits a solution. The relaxation of the entropy-minimization problem however generally engenders the loss of a one-to-one correspondence between the moments and the distribution. Moreover, relaxation of the entropy minimization problem does not resolve the potential singularities in the flux function, as these singularities are intrinsic to the exponential form of the closure relation. Another fundamental complication, pertaining to the implementation of moment systems based on exponential closure, is that the resulting formulation requires the evaluation of moments of exponentials of polynomials of, in principle, arbitrary order. It is generally accepted that the derivation of closed-form expressions for such moments is intractable, and 
accurate approximation of the moments is a notoriously difficult problem; see, for instance,~\cite{Lasserre2010}. 

In this paper we consider alternative moment-closure relations for the Boltzmann equation, based on approximations of the exponential function 
derived from truncations of its standard limit definition, $\exp(\cdot)=\lim_{n\to\infty}(1+\cdot/n)^n$.
It is to be noted that closure relations derived from a series-expansion definition of the exponential have received scant attention before, e.g., by Brini and Ruggeri~\cite{Brini:2002kx}. Our motivation for considering the limit definition instead of the series-expansion definition for constructing the moment closures
is based on the direct availability of a corresponding inverse relation for higher order approximations. We propose a generalization of the setting of the moment-closure problem from Kullback--Leibler divergence~\cite{Kullback1951} (i.e relative entropy) to 
the class of $\varphi$\nobreakdash-divergences \cite{Csiszar1972}. The considered $\varphi$\nobreakdash-divergences constitute an approximation to the Kullback--Leibler divergence in the vicinity of some Maxwellian. It will be shown that the approximate-exponential closure relation can be derived
via constrained minimization of a corresponding $\varphi$\nobreakdash-divergence. The proposed description encapsulates as special cases Grad's 
closure relation and Levermore's entropy-based closure relation. For even order approximations of the exponential, the closure relation engenders 
non-negative phase-space distributions. Moreover, the corresponding moment systems are symmetric hyperbolic and tractable, in the sense that 
the formulation only requires the evaluation of higher-order moments of Gaussian distributions. The moment systems furthermore dissipate 
an appropriate $\varphi$-divergence, analogous to the dissipation of relative entropy of the Boltzmann equation, provided that the collision operator
dissipates the corresponding $\varphi$-divergence. We will show that the class of collision operators that dissipate appropriate $\varphi$-divergences 
includes the standard BGK~\cite{Bhatnagar:1954hc} and generalized BGK~\cite{Levermore1996} operators.

The remainder of this paper is organized as follows. Section \ref{sec:BoltzProps} abstracts, for completeness, well known structural features of the Boltzmann equation to be retained in the developed moment system approximation. Section \ref{sec:MomSys} introduces concepts relevant to moment systems pertaining to subspace approximations and reviews the moment closures of Grad \cite{Grad1949} and Levermore \cite{Levermore1996} in light of the aforementioned issues, namely, admissibility of phase-space distributions, hyperbolicity, realizability and tractability. Section \ref{sec:DivMomCls} presents a novel tractable moment closure approximation and, moreover, it will be shown that the corresponding closed system of moment equations are well-posed and retain the structural features of the Boltzmann equation. Finally, section \ref{sec:Conc} gives a concluding discussion.

\section{The Boltzmann Equation}
\label{sec:BoltzProps}
Consider a monatomic gas, i.e.  a gas composed of a single species of identical classical particles, contained in a fixed spatial
domain $\Omega\subset\mathbb{R}^D $. Kinetic theory describes the state of such a gas by a non-negative (phase-space) 
density $f=f(t,\bm x,\bm v)$ over the single-particle phase space $\Omega\times\mathbb{R}^D$. The evolution of $f$ is considered 
to be governed by the Boltzmann equation,
\begin{align}\label{eq:Boltzmann}
\partial_t f + v_i\partial_{x_i} f=\CC(f)
\end{align}
where the collision operator $f\mapsto\CC(f)$ acts only on the $\bm v=(v_1,\ldots,v_D)$ dependence  of $f$ locally at each $(t,\bm x)$
and the summation convention applies to repeated indices. 
The collision operator is assumed to possess certain conservation, symmetry and dissipation properties, viz.,
conservation of mass, momentum and energy, invariance under Galilean transformations and dissipation of appropriate entropy functionals. 
These fundamental properties of the collision operator are treated in further detail below. Our treatment of the conservation
and symmetry properties is standard (see, for instance,~\cite{Levermore1996}) and is presented merely for coherence and completeness. 
For the entropy-dissipation property, we consider a generalization of the usual (relative) entropy to $\varphi$-divergences~\cite{Csiszar1972},
to enable an exploration of the moment-closure problem in an extended setting; see Section~\ref{sec:DivMomCls}.

To elaborate the conservation properties of the collision operator, 
let $\langle \cdot \rangle$ denote integration in the velocity dependence of any scalar, vector or matrix valued measurable function 
over $D$\nobreakdash-dimensional Lebesgue measure. A function $\psi:\IR^D\to\IR$ is called a {\em collision invariant\/} of $\CC$
if
\begin{align}
\label{eq:DefCollInvariant}
\langle \psi\,\CC(f) \rangle = 0  \qquad \forall f\in \Domain(\CC),
\end{align}
where $\Domain(\CC)\subset{}L^1(\IR^D,\IR_{\geq{}0})$ denotes the domain of~$\CC$, which we consider to be a subset of the almost everywhere nonnegative
Lebesgue integrable functions on~$\IR^D$. 
Equation~\EQ{Boltzmann} associates a scalar conservation law with each collision invariant:
\begin{equation}
\label{eq:ConsvLaw}
\partial_t\langle \psi f\rangle+\partial_{x_i} \langle v_i \psi f \rangle = 0
\end{equation}
We insist that $\{1,v_1,\ldots,v_D,|\bm v|^2\}$ are collision invariants of $\CC$ and that 
the span of this set contains all collision invariants, i.e.
\begin{equation*}\label{eq:CollInvariant}
\langle{}\psi\,\CC(f)\rangle=0\quad\forall{}f\in\Domain(\CC)
\quad\Leftrightarrow\quad
\psi\in\mathrm{span}\{1,v_1,\ldots,v_D,|\bm v|^2\}=:\IE.
\end{equation*} 
The moments $\langle{}f\rangle$,
$\langle{}v_if\rangle$ and $\langle{}|{\bm v}|^2f\rangle$, correspond to mass-density, the (components of) momentum-density and energy-density, respectively.
Accordingly, the conservation law~\EQ{ConsvLaw} implies that~\EQ{Boltzmann} conserves mass, momentum and energy.

The assumed symmetry properties of the collision operator pertain to commutation with translational and rotational transformations. 
In particular, for all vectors $\bm u\in\mathbb{R}^D$ and all orthogonal tensors $\mathcal{O}:\IR^D\to\IR^D$, we define the translation transformation
$\mathcal{T}_{\bm u}:\Domain(\CC)\to\Domain(\CC)$ and the rotation transformation $\mathcal{T}_{\mathcal{O}}:\Domain(\CC)\to\Domain(\CC)$ by:
\begin{alignat*}{2}
(\mathcal{T}_{\bm u}f)(\bm v)&=f(\bm u-\bm v)&\qquad&\forall{}f\in\Domain(\CC)
\\
(\mathcal{T}_{\mathcal{O}}f)(\bm v)&=f(\mathcal{O}^*\bm v) &\qquad&\forall{}f\in\Domain(\CC)
\end{alignat*}
with $\mathcal{O}^*$ the Euclidean adjoint of $\mathcal{O}$. Note that the above transformations act on the $\bm v$-dependence only. 
It is assumed that $\CC$ possesses the following symmetries:
\begin{equation}
\label{eq:GalilInvar}
\CC(\mathcal{T}_{\bm u}f)=\mathcal{T}_{\bm u}\CC(f),\qquad
\CC(\mathcal{T}_{\mathcal{O}}f)=\mathcal{T}_{\mathcal{O}}\CC(f)
\end{equation}
The symmetries (\ref{eq:GalilInvar}) imply that (\ref{eq:Boltzmann}) complies with Galilean invariance, i.e.
if $f(t,\bm x,\bm v)$ satisfies the Boltzmann equation~\EQ{Boltzmann}, then for arbitrary ${\bm u}\in\IR^D$ and arbitrary orthogonal $\mathcal{O}:\IR^D\to\IR^D$, so 
do $f(t,\bm x - {\bm u}t,{\bm v}-{\bm u})$ and $f(t,\mathcal{O}^*{\bm x},\mathcal{O}^*{\bm v})$.

The entropy dissipation property of~$\CC$ is considered in the extended setting of~\cite[Sec.~7]{Levermore1996}, from which we 
derive the following definition: a convex function $\HH:\IR_{\gneq{}0}\to\IR$ is called an {\em entropy density for $\CC$\/} if
\begin{equation}
\label{eq:Dissipation}
\langle \HH'(f)\,\CC(f) \rangle \leq 0, \qquad \forall f\in\Domain(\CC)
\end{equation}
with $\HH'(f)$ the derivative of $\HH(f)$, and if for every $f\in\Domain(\CC)$ the following equivalences hold:
\begin{equation}
\label{eq:Equilibrium}
\CC(f) = 0
\quad\Leftrightarrow\quad
\langle \HH'(f)\,\CC(f)\rangle  = 0
\quad\Leftrightarrow\quad
\HH'(f)\in\IE
\end{equation}
Relation (\ref{eq:Dissipation}) implies that~$\CC$ dissipates the local entropy $\langle\HH(\cdot)\rangle$, which leads to an
abstraction of Boltzmann's H-theorem for~\EQ{Boltzmann}, asserting that solutions of the Boltzmann equation (\ref{eq:Boltzmann}) satisfy the 
local entropy-dissipation law:
\begin{align}
\label{eq:EntDiss}
 \partial_t\langle \HH(f) \rangle+\partial_{x_i} \langle v_i \HH(f) \rangle = \langle \CC(f)\, \HH'(f)\rangle \leq 0\,.
\end{align}
The functions $\langle \HH(f) \rangle$,  $\langle v_i \HH(f) \rangle$ and $\langle \HH'(f)\, \CC(f)\rangle$ are referred to as entropy density, entropy flux and entropy-dissipation rate, respectively.
The first equivalence in~(\ref{eq:Equilibrium}) characterizes local equilibria of~$\CC$ 
by vanishing entropy dissipation, while the second equivalence indicates the form of such local equilibria.
For spatially homogeneous initial data, $f_0$, Equations~\EQ{Dissipation} and~\EQ{Equilibrium} suggest that equilibrium solutions, $\feq$, 
of~\EQ{Boltzmann} are determined by:
\begin{equation}
\label{eq:LegEq}
\feq=\argmin\big\{\langle\HH(f)\rangle:f\in\Domain(\CC),\langle f \bm\psi\rangle=\langle{}f_0\bm\psi\rangle\},
\end{equation}
Equation~\EQ{LegEq} identifies equilibria as minimizers\footnote{We adopt the sign convention of diminishing entropy.} of the entropy, subject to the constraint that the invariant moments are identical to the invariant moments of the initial distribution. 

The standard definition of entropy corresponds to a density $f\mapsto{}f\log{}f$, possibly augmented with~$f\psi$ where $\psi\in\IE$ is any collision invariant.
It is to be noted that for Maxwellians $\MM$, i.e. distributions of the form
\begin{equation} \label{eq:Maxwellian}
\MM({\bm v}):=
\MM_{(\varrho,{\bm u},T)}({\bm v}) :=
\frac{\varrho}{(2\pi{}RT)^{\frac{D}{2}}}\exp\left(-\frac{|\bm v-\bm u|^2}{2RT}\right)
\end{equation}
for some $(\varrho, {\bm u}, T)\in\mathbb{R}_{>0}\times\mathbb{R}^D\times\mathbb{R}_{>0}$ and a certain gas constant $R\in\IR_{>0}$, it holds that $\log\MM\in\IE$. Therefore, the relative entropy $\langle{}f\log{}(f/\MM)\rangle$ of~$f$ with  respect to~$\MM$ is equivalent to $\langle{}f\log{}f\rangle$ in the sense of dissipation characteristics.
The physical interpretation of the entropy $\langle{}f\log{}f\rangle$, due to Boltzmann~\cite{Boltzmann1868,Boltzmann1877,Boltzmann2011}, is that 
of a measure of degeneracy of macroscopic states, i.e. of the number of microscopic states that are consistent with the macroscopic state as described 
by the one-particle marginal, $f$. In the context of information theory, Shannon~\cite{Shannon1948} showed that for discrete probability distributions, the density $f\mapsto{}f\log{}f$ is uniquely defined by the postulates of continuity, strong additivity and the property that $m\HH(1/m)<n\HH(1/n)$ whenever $n<m$. These postulates ensure that for discrete probability distributions the entropy yields a meaningful characterization of information content and, accordingly, rationalize an interpretation of entropy as a measure of the uncertainty or, conversely, information gain pertaining to an observation represented by the corresponding probability distribution~\cite{Jaynes1957}. Kullback and Leibler \cite{Kullback1951} generalized Shannon's definition of information to the abstract case
and identified the divergence%
\footnote{The conventional definition of Kullback--Leibler divergence according to~\EQ{KLDiv} is historically incorrect,
as Kullback and Leibler in fact referred to the symmetrization of~\EQ{KLDiv} as the ``divergence''.}
\begin{equation}
\label{eq:KLDiv}
D_{\text{KL}}(\mu_1|\mu_2)=\int f_1\log(f_1/f_2)\,d\nu
\end{equation}
as a distance between mutually absolutely continuous measures $\mu_1$ and~$\mu_2$, both absolutely continuous with respect to the measure $\nu$
with Radon--Nikodym derivatives $f_1=d\mu_1/d\nu$ and $f_2=d\mu_2/d\nu$. The Kullback--Leibler divergence characterizes the
mean information for discrimination between $\mu_1$ and~$\mu_2$ per observation from $\mu_1$. Noting that the Kullback--Leibler divergence~\EQ{KLDiv}
coincides with the relative entropy of $f_1$ with respect to~$f_2$, the relative entropy $\langle{}f\log(f/\MM)\rangle$ can thus be understood 
as a particular measure of the divergence of the one-particle marginal relative to the reference (or {\em background\/}) distribution~$\MM$.
Kullback--Leibler divergence was further generalized by Csisz\'{a}r~\cite{Csiszar1972} and Ali \textit{et. al.}~\cite{Ali1966}, who introduced 
a general class of distances between probability measures, referred to as $\varphi$-divergences, of the form:
\begin{equation}
\label{eq:PhiDiv}
D_{\varphi}(\mu_1|\mu_2)=\int f_2\,\varphi({f_1}/{f_2})\,d\nu
\end{equation}
where $\varphi$ is some convex function subject to $\varphi(1)=\varphi'(1)=0$ and $\varphi''(1)>0$. Note that the Kullback--Leibler divergence corresponds to the specific case $\varphi_{\text{KL}}(\cdot)=(\cdot)\log(\cdot)$.

In this work, we depart from the standard (relative) entropy for~\EQ{Boltzmann} and instead consider entropies based on particular
$\varphi$\nobreakdash-divergences. These $\varphi$\nobreakdash-divergences generally preclude the usual physical and information-theoretical 
interpretations, but still provide a meaningful entropy density in accordance with~(\ref{eq:Dissipation}) and~(\ref{eq:Equilibrium}). 
The considered $\varphi$\nobreakdash-divergences yield a setting in which entropy-minimization based moment-closure approximations to~\EQ{Boltzmann} 
are not impaired by non-realizability, exhibit bounded fluxes in the vicinity of equilibrium, and are numerically tractable.

\begin{remark} 
\label{rem:remark1}
Implicit to our adoption of $\varphi$-divergence-based entropies is the assumption that such entropies comply with~\EQ{Dissipation} and~\EQ{Equilibrium} for a 
meaningful class of collision operators. It can be shown that the class of admissible collision operators includes the BGK operator~\cite{Bhatnagar:1954hc}:
\begin{equation}
\label{eq:BGK}
\CCBGK(f) = -\tau^{-1}(f-\mathcal{E}_f)
\end{equation}
where $\tau\in\IR_{>0}$ is a relaxation time and $\mathcal{E}_{(\cdot)}$ corresponds to the map $f_0\mapsto\feq$ defined by~\EQ{LegEq}.
The Kuhn--Tucker optimality conditions associated with~\EQ{LegEq} convey that $\HH'(\mathcal{E}_f)\in\IE$ and, therefore,
$\langle{}\HH'(\mathcal{E}_f)(f-\mathcal{E}_f)\rangle=0$. The dissipation inequality~\EQ{Dissipation} then follows from the
convexity of~$\eta(\cdot)$:
\begin{equation}
\label{eq:MonInc}
\langle \HH'(f)\, \CCBGK(f) \rangle
=
-\tau^{-1}\big\langle\big(\HH'(f)-\HH'(\mathcal{E}_f)\big)(f-\mathcal{E}_f)\big\rangle 
\leq0
\end{equation}
Moreover, because equality in~\EQ{MonInc} holds if and only if $f=\mathcal{E}_f$, the condition $\langle\HH'(f)\,\CCBGK(f)\rangle=0$ implies
that $f=\mathcal{E}_f$, which in turn yields $\CCBGK(f)=0$ and $\HH'(f)\in\IE$. The equivalences in~\EQ{Equilibrium} are therefore also verified. A similar result holds for the multi-scale generalization of the BGK operator introduced in~\cite{Levermore1996}; see Appendix \ref{Appendix:GBGK}.
\end{remark}

\section{Moment Systems}
\label{sec:MomSys}
Moment systems are approximations of the Boltzmann equation based on a finite number of velocity-moments of the one-particle marginal. An inherit aspect of moment equations derived from~\EQ{Boltzmann} is that low-order moments are generally coupled with higher-order ones, and consequently a closed set of equations for the moments cannot be readily formulated. Therefore, a closure relation is required.

To derive the moment equations from~\EQ{Boltzmann} and elaborate on the corresponding moment-closure problem, 
let $\mathbb{M}$ denote a finite-dimensional subspace of $D$-variate polynomials and 
let $\{m_i(\bm v)\}_{i=1}^M$ represent a corresponding basis. Denoting the column $M$\nobreakdash-vector of these 
basis elements by $\bm m$, it holds that the moments $\{\langle{}{m_i}f\rangle\}_{i=1}^M$ of the one-particle marginal satisfy:
\begin{align} \label{eq:MomSys}
\partial_t\langle \bm m  f\rangle+\partial_{x_i}\langle v_i\bm m  f\rangle = \langle \bm m\mathcal{C}(f)\rangle
\end{align}
It is to be noted that we implicitly assume in~\EQ{MomSys} that $f$ resides in
\begin{equation}
\label{eq:IF}
\IF\defeq\big\{f\in{}\Domain(\CC):mf\in{L}^1(\mathbb{R}^{D}),\,\bm v mf\in{L}^1(\mathbb{R}^{D},\IR^D),\,m\CC(f)\in{}L^1(\IR^D)\text{ for all }m\in\mathbb{M}\big\}
\end{equation}
almost everywhere in the considered time interval $(0,T)$ and the spatial domain $\Omega$.
This assumption has been confirmed in specific settings of~\EQ{Boltzmann} but not for the general case; see~\cite[Sec. 4]{Levermore1996} and
the references therein for further details. The moment-closure problem pertains to the fact that~\EQ{MomSys}
provides only $M$ relations between $(2+D)M$ independent variables, viz., the densities $\langle{} m_i f\rangle$, the flux components 
$\langle{}v_i m_i f\rangle$ and the production terms $\langle m_i\CC(f)\rangle$. Therefore, $(1+D)M$ auxiliary relations must be
specified to close the system. Generally, moment systems are closed by expressing the fluxes and production terms as
a function of the densities. 
Moment systems are generally closed by constructing an approximation to the distribution function from the densities and
then evaluating the fluxes and production terms for the approximate distribution. Denoting by
$\IA\subseteq\IR^M$ a suitable class of moments, a function $\ff:\IA\to\IF$ must be
specified such that~$\ff$ realizes the moments in~$\IA$, i.e. $\langle\bm m\ff(\bm\rho)\rangle=\bm\rho$ for all $\bm\rho\in\IA$,
and $\ff(\langle{}\bm m f\rangle)$ constitutes a suitable (in a sense to be made more precise below) approximation 
to the solution~$f$ of the Boltzmann equation~\EQ{Boltzmann}. 
Approximating the moments in~\EQ{MomSys} by $\bm\rho\approx\langle\bm m f\rangle$ and replacing $f$ in~\EQ{MomSys} by the 
approximation $\ff(\bm\rho)$, one obtains the following closed system for the approximate moments:
\begin{equation} 
\label{eq:ClsMomSys}
\partial_t\bm \rho+\partial_{x_i}\langle v_i\bm m  \mathcal{F}(\bm\rho)\rangle=\langle\bm m \mathcal{C}(\mathcal{F}(\bm\rho))\rangle.
\end{equation}
The closed moment system~\EQ{ClsMomSys} is essentially defined by the polynomial subspace, $\IM$, and the closure relation,~$\ff$.
A subspace/closure-relation pair $(\IM,\ff)$ is suitable if the corresponding moment system~(\ref{eq:ClsMomSys}) is well posed
and retains the fundamental structural properties of the Boltzmann equation~(\ref{eq:Boltzmann}) as described in section \ref{sec:BoltzProps}, viz.,
conservation of mass, momentum and energy, Galilean invariance and dissipation of an entropy functional. Auxiliary conditions
may be taken into consideration, e.g. that the fluxes and production terms can be efficiently evaluated by numerical quadrature.

It is noteworthy that moment systems can alternatively be conceived of as Galerkin subspace-approximations of the Boltzmann equation in renormalized form. This Galerkin-approximation interpretation can for instance
prove useful in constructing error estimates for~\EQ{ClsMomSys} and in deriving structural properties.
In addition, the Galerkin-approximation interpretation conveys that smooth functionals
of approximate distributions obtained from moment systems, such as velocity moments, 
generally display {\em superconvergence\/} under hierarchical-rank refinement, in accordance with the Babu\v{s}ka--Miller theorem; see~\cite{Babuska:1984fk} and also Section~\ref{sec:NumExp}.
Consider the subspace $\IM$ and let $\beta:\IM\to\IF$ denote a renormalization map.
Denoting by $V((0,T)\times\Omega;\IM)$ a suitable class of functions from~$(0,T)\times\Omega$ into~$\IM$, the moment system~\EQ{ClsMomSys} can be recast into the Galerkin form:
\begin{multline}
\label{eq:GalLevMomCls}
\text{\it Find }
g\in{}V\big((0,T)\times\Omega;\IM)\big):\\
\big\langle m\partial_t \beta(g)\big\rangle+\big\langle mv_i\partial_{x_i}\beta(g)\big\rangle = \big\langle m\mathcal{C}(\beta{}(g))\big\rangle
\quad
\forall m\in{}\IM,\text{ a.e. }(t,\bm x)\in(0,T)\times\Omega.
\end{multline}
To elucidate the relation between~\EQ{ClsMomSys} and~\EQ{GalLevMomCls}, we associate to $\beta:\IM\to\IF$ a function $\ff_{\beta}:\Domain(\ff_{\beta})\to\IF$ such that $\ff_{\beta}(\bm\rho)=\beta{}(g_{\bm\rho})$ with $g_{\bm\rho}$ according to $\langle\bm m\beta{}(g_{\bm\rho})\rangle=\bm\rho$. The domain $\Domain(\ff_{\beta})$ is implicitly
restricted to moments $\bm\rho\in\IR^M$ that can be realized by some $g\in\IM$. The equivalence between the Galerkin formulation~\EQ{GalLevMomCls} and the moment system~\EQ{ClsMomSys} now follows immediately by noting that $\{m_i\}_{i=1}^M$ constitutes a basis of~$\IM$ and inserting $g_{\bm\rho}$ for $g$ in~\EQ{GalLevMomCls}.

In the remainder of this section we review the celebrated moment closures of Levermore~\cite{Levermore1996} and Grad~\cite{Grad1949} to provide a basis for the subsequent divergence-based moment closures in section~\ref{sec:DivMomCls}.

\subsection{Levermore's Entropy-Based Moment Closure}
\label{sec:LevMomCls}
The moment-closure relation of Levermore~\cite{Levermore1996} is essentially characterized by the renormalization map~$\beta(\cdot)=\exp(\cdot)$. For this closure relation, 
a subspace $\mathbb{M}$ is considered to be admissible if it satisfies:
%
%
\begin{enumerate}
\renewcommand{\theenumi}{\arabic{enumi}}
\renewcommand{\labelenumi}{\theenumi)}
\item $\IE \subseteq \mathbb{M}$;
\item $\mathbb{M}$ is invariant under the actions of $\mathcal{T}_{\bm u}$ and
$\mathcal{T}_{\mathcal{O}}$;
\item $\mathbb{M}_c:=\{m\in\mathbb{M}: \left\langle \exp(m)\right\rangle<\infty\}$ has a nonempty interior in~$\mathbb{M}$.
\end{enumerate}
The first condition insists that $\IM$ contains the collision invariants, which ensures that the moment system imposes conservation of mass, momentum and energy. 
These conservation laws must be obeyed if any fluid-dynamical approximation is to be recovered. The second condition dictates that for all $m\in\IM$, all $\bm u\in\IR^D$ and
all orthogonal tensors~$\mathcal{O}$ it holds that
$m(\bm u-(\cdot))\in\IM$ and $m(\mathcal{O}^*(\cdot))\in\IM$. This condition ensures that
the moment system exhibits Galilean invariance. As argued by Junk~\cite{Junk2002}, rotation and translation invariant finite dimensional spaces 
are necessarily composed of multivariate polynomials. 
The third condition requires that~$\mathbb{M}$ contains functions~$m$ such that $\beta(m(\cdot))$ is Lebesgue integrable on~$\IR^D$.
For $\beta(\cdot)=\exp(\cdot)$ and $\IM$ composed of multivariate polynomials, this condition implies that the highest-order terms in any variable in~$\IM$ must be of 
even order. The subset $\IM_c$ then corresponds to a convex cone, consisting of all polynomials in~$\IM$ for which the highest-order terms in any variable are of even order and have a negative coefficient. One can infer that $\exp(\cdot)$ maps~$\IM_c$ to distributions with bounded moments and fluxes, i.e. $g\in\IM_c$ implies 
$|\langle{}m\beta(g)\rangle|<\infty$ and $|{\bm v}m\beta(g)\rangle|<\infty$ for all~$m\in\IM$.

In~\cite{Levermore1996} the moment-closure relation associated with $\beta(\cdot)=\exp(\cdot)$ is derived by minimization of the entropy with density 
$\eta_{\mathrm{L}}(f):=f\log{}f-f$, subject to the moment constraint. Specifically, considering any admissible subspace~$\IM$, Levermore formally 
defines the closure relation $\bm\rho\mapsto\FL(\bm\rho)$ according to:
\begin{equation}
\label{eq:EntMin}
\FL(\bm\rho):=
\argmin_{f\in\mathbb{F}}
\big\{\langle f\log f - f\rangle:\langle \bm m f \rangle=\bm\rho\big\}
\end{equation}
To elucidate the fundamental properties of the closure relation~\EQ{EntMin}, we consider an admissible subspace~$\IM$ and we denote by $\ID$ the collection
of all~$f\in\IF$ that yield moments $\bm\rho=\langle\bm m f\rangle$ for which the minimizer in~\EQ{EntMin} exists. The operator
$\FL(\langle\bm m(\cdot)\rangle):\ID\to\II$ is idempotent and its image~$\II\subset\ID$ admits a finite-dimensional characterization. In particular, 
it holds that~$\log\II$ coincides with the convex cone~$\IM_c$. The idempotence of the operator~$\FL(\langle\bm m(\cdot)\rangle)$ and its injectivity in~$\ID$ imply 
that~\EQ{EntMin} corresponds to a projection. This projection is generally referred to as the {\em entropic projection\/}~\cite{Hauck2006}. 
A second characterization of~\EQ{EntMin} follows from the following sequence of identities, which holds for any $f\in\IF$ such that $\langle\bm m f\rangle={\bm\rho}$
and all Maxwellian distributions, $\MM=\exp(\psi)$ with $\psi\in\IE$:
\begin{equation}
\label{eq:SeqID1}
\langle{}f\log(f/\MM)\rangle
=
\langle{}f(\log{}f-\psi)\rangle
=
\langle{}f\log{}f-f\rangle+\langle{}f(1-\psi)\rangle
=
\langle{}f\log{}f-f\rangle+{\bm\alpha}\cdot{\bm\rho}
\end{equation}
for some $\bm\alpha\in\IR^M$. Noting that $\bm\alpha\cdot\bm\rho$ is independent of~$f$, one can infer from~\EQ{KLDiv} and~\EQ{SeqID1} that~$\FL$ 
according to~\EQ{EntMin} is the distribution in~$\IF$ that is closest to equilibrium in the Kullback--Leibler divergence, subject to the condition that its 
moments~$\langle\bm m (\cdot)\rangle$ coincide with~$\bm\rho$. Similarly, it can be shown that $\FL$ according to~\EQ{EntMin} minimizes $\langle{}f\log{}f\rangle$
subject to $\langle\bm m f\rangle=\bm\rho$. Therefore, the information interpretation of the entropy $\langle{}f\log{}f\rangle$ (see Section~\SEC{BoltzProps})
enables a third characterization of~\EQ{EntMin}, viz. as the least-biased distribution given the information $\langle\bm m(\cdot)\rangle=\bm\rho$ on the moments.  

The exponential form of the renormalization map associated with~\EQ{EntMin} can be derived straightforwardly by means of the Lagrange multiplier method.
Provided it exists, the minimizer of the constrained minimization problem~\EQ{EntMin} corresponds to a stationary point of the Lagrangian 
$(f,\bm\alpha)\mapsto\langle{}f\log{}f-f{}\rangle+\bm\alpha\cdot(\bm\rho-\langle{}\bm m f\rangle)$. The stationarity condition implies that
$\log{}f-\bm\alpha\cdot\bm m$ vanishes, which conveys the exponential form $f=\exp(\bm\alpha\cdot\bm m)$. It is to be noted that the Lagrange multipliers
have to comply with an admissibility condition related to integrability. In particular, $\bm\alpha\cdot\bm m$ must belong to the convex cone~$\IM_c$.

In~\cite{Levermore1996} it is shown that the moment system~\EQ{ClsMomSys} with closure $\FL$ corresponds to a quasi-linear symmetric hyperbolic system for the Lagrange multipliers. Application of the chain rule to~\EQ{ClsMomSys} with $\FL(\bm\rho)=\exp(\bm\alpha\cdot\bm m)$ (with, implicitly, $\bm\rho=\langle\bm{m}\exp(\bm\alpha\cdot\bm{m})\rangle$) yields:
\begin{equation}
\label{eq:HypSys}
{\bm A}_0(\bm\alpha)\frac{\partial{\bm\alpha}}{\partial{}t}
+
\sum_{i=1}^D{\bm A}_i(\bm\alpha)\frac{\partial\bm\alpha}{\partial{}x_i}
=
{\bm s}(\bm\alpha)
\end{equation}
with ${\bm A}_0(\bm\alpha)=\langle\bm m \otimes \bm m \exp(\bm\alpha\cdot\bm m)\rangle$,
${\bm A}_i(\bm\alpha)=\langle v_i\bm m \otimes \bm m \exp(\bm\alpha\cdot\bm m)\rangle$ and
${\bm s}(\bm\alpha)= \langle \bm m \, \CC(\exp(\bm\alpha\cdot\bm m))\rangle$.
The symmetry of~${\bm A}_i$ $(i=0,1,\ldots,D)$ and the positive definiteness of ${\bm A}_0$ are evident. 
By virtue of its quasi-linear symmetric hyperbolicity, the system~\EQ{HypSys} is at least linearly well posed~\cite{Levermore1996}. 
Moreover, under auxiliary conditions on the initial data, local-in-time existence of solutions can be established; see, for instance, \cite{Majda:1984wj}. 

Levermore's moment systems retain the fundamental structural properties of the Boltzmann equation. The conservation properties and Galilean invariance are direct consequences of conditions~1. and~2. on the admissible subspaces, respectively. Dissipation of the entropy $\langle{}\eta_{\mathrm{L}}(\cdot)\rangle$ can be inferred from
the Galerkin formulation~\EQ{GalLevMomCls}, by noting that for $\beta(\cdot)=\exp(\cdot)$ it holds that $\log\beta(\cdot):\IM\to\IM$. Hence, if $g$ complies with~\EQ{GalLevMomCls} and $\beta(\cdot)=\exp(\cdot)$ then the following identity holds on account of Galerkin orthogonality:
\begin{equation}
\langle{}\log\beta(g)\,\partial_t\beta(g)\rangle+\langle{}\log\beta(g)\,v_i\partial_{x_i}\beta(g)\rangle=\langle\log\beta(g)\,\CC(\beta(g))\rangle
\end{equation}
The left-hand side of this identity coincides with $\partial_t\langle\eta_{\mathrm{L}}(\beta(g))\rangle+\partial_{x_i}\langle{}v_i\eta_{\mathrm{L}}(\beta(g))\rangle$, while
the right-hand side equals $\langle{}\CC(\beta(g))\,\eta_{\mathrm{L}}'(\beta(g))\rangle$. For~$g$ according to~\EQ{GalLevMomCls}, the distribution~$\beta(g)=\exp(g)$ thus obeys the entropy dissipation relation~\EQ{EntDiss} with entropy density~$\eta_{\mathrm{L}}$.

Levermore's consideration of entropy-based moment-closure systems in~\cite{Levermore1996}, as well as the above exposition,
implicitly rely on existence of a solution to the moment-constrained entropy minimization problem~\EQ{EntMin}. It was however shown by
Junk in the series of papers~\cite{Junk1998,Junk2000,Junk2002} that for super-quadratic~$\IM$ the closure relation~\EQ{EntMin} is impaired 
by non-realizability, i.e. a minimizer of~\EQ{EntMin} may be non-existent. Moreover, the class of local equilibrium distributions generally lies on the boundary of the set of degenerate densities. In~\cite{Junk1998}, Junk also establishes that the flux $\langle v_i\bm m \beta(g)\rangle$ 
can become unbounded in the vicinity of equilibrium, thus compromising well-posedness of~\EQ{HypSys}.
The singularity of the fluxes moreover represents a severe complication for numerical approximation methods; see also~\cite{McDonald:2013xu}.

The realizability problem of Levermore's entropy-based moment closure has been extensively investigated; see, in 
particular,~\cite{Junk1998,Junk2000,Junk2002,Hauck2008,Schneider2004,Pavan2011}. In~\cite{Junk2002,Hauck2008,Pavan2011} it has been
shown that the set of degenerate densities is empty if and only if the set 
$\{\bm{\alpha}\in\IR^M:\bm m \exp(\bm\alpha\cdot\bm m)\in{}L^1(\IR^D,\IR^M)\}$
of Lagrange multipliers associated with integrable distributions is open. This result implies that
degenerate densities are unavoidable for super-quadratic polynomial spaces, because the Lagrange
multipliers associated with equilibrium are then located on the boundary of the above set; see also~\cite{Hauck2008}. 
To bypass the realizability problem, Schneider~\cite{Schneider2004} and Pavan~\cite{Pavan2011} considered the
following relaxation of the constrained entropy-minimization problem:
\begin{equation}
\label{eq:2.15}
\argmin_{f\in\IF}\{\langle{}f\log{}f-f\rangle:\langle \bm m  f\rangle\leq^*\bm\rho\}
\end{equation}
where the binary relation $\leq^*$ connotes that the highest order moments of the left member are
bounded by the corresponding moments of the right member. The relaxation of the highest-order-moment
constraints serves to accommodate that minimizing sequences $\{f_n\}\subset\IF$ subject to the 
constraint $\langle{}\bm m f_n\rangle=\bm\rho$ converge (in the topology of absolutely integrable functions)
to an exponential density with inferior highest-order moments; see~\cite{Junk1998,Junk2000,Schneider2004,Hauck2008,Pavan2011}.
The analyses in~\cite{Schneider2004,Pavan2011} convey that the relaxed minimization problem indeed admits a unique
solution, corresponding to an exponential distribution. The exponential closure can therefore be retained if the closure relation 
is defined by~\EQ{2.15} instead of~\EQ{EntMin}. It is to be noted however that the closure relation~\EQ{2.15}
does not generally provide a bijection between the Lagrange multipliers and the moments. Moreover, the aforementioned 
singularity of fluxes near equilibrium is also inherent to~\EQ{2.15}.

Another formidable obstruction to the implementation of numerical approximations of Levermore's moment-closure systems are
the exponential integrals that appear in~(\ref{eq:ClsMomSys}). The evaluation of moments of exponentials of super-quadratic 
polynomials is generally accepted to be intractable, and accurate approximation of such moments is algorithmically complicated
and computationally intensive; see, in particular, \cite[Sec.~12.2]{Lasserre2010} and~\cite[Sec.~6]{Junk1998}. 

\subsection{Grad's Hermite-Based Moment Closure}
\label{sec:Grad}
In his seminal paper~\cite{Grad1949}, Grad proposed a moment-closure relation based on a
factorization of the one-particle marginal in a Maxwellian distribution and a term
expanded in Hermite polynomials; see
also~\cite[Sec.~V]{Grad:1958ek}.
%
%
The expansion considered by Grad writes:
\begin{equation}
\label{eq:GradDist}
f(t,\bm{x},\bm{c})\approx\mathcal{M}(\bm c)
\sum_{k=0}^{n}\sum_{\bm{i}_k}\frac{1}{k!}a_{\bm{i}_k}^{(k)}(\bm{x},t)\mathscr{H}_{\bm{i}_k}^{(k)}(\bm{c}),
\end{equation}
where $\bm c$ denotes peculiar velocity, $\bm{i}_k=(i_1,i_2,\ldots,i_k)$ is a multi-index with sub-indices $\smash[b]{i_{(\cdot)}}\in\{1,2,\ldots,D\}$, $\smash[t]{\smash[b]{a_{\bm{i}_k}^{(k)}}}$ are the polynomial expansion coefficients and $\smash[t]{\smash[b]{\mathscr{H}{}_{\bm{i}_k}^{(k)}}}$ are $D$-variate Hermite polynomials of degree~$k$:
\begin{equation}
\label{eq:Hermite}
\mathscr{H}{}^{(k)}_{\bm{i}_k}(\bm{x})
=
\frac{(-1)^k}{\omega(|\bm{x}|)}\frac{\partial^k\omega(|\bm{x}|)}{\partial{}x_{i_1}\partial{}x_{i_2}\cdots\partial{}x_{i_k}}
\quad\text{with}\quad
\omega(s)=\frac{1}{(2\pi)^{d/2}}\exp(-s^2/2).
\end{equation}
The Maxwellian in~\EQ{GradDist} can either correspond to a prescribed local or global Maxwellian, or it can form part of the approximation;
see~\cite{Grad1949,Grad:1958ek}. In the latter case, the coefficients associated with invariant moments are fixed and it holds that
$\smash[t]{a^{(0)}=1}$, $\smash[t]{a^{(1)}_i=0}$ and~$\smash[t]{a^{(2)}_{ii}=1}$. 
By virtue of the specific properties of Hermite polynomials, moments of 
Grad's approximate distribution~\EQ{GradDist} can be evaluated in closed-form. 

The linear hull of the Hermite polynomials $\smash[t]{\{\mathscr{H}^{(k)}_{\bm i}\}_{0\leq{}k\leq{}n}}$ coincides with the class of $D$-variate polynomials of degree at most~$n$.
The Hermite polynomials in~\EQ{Hermite} do not provide a basis of the polynomials, however, on account of linear dependence; evidently, the Hermite polynomial in~\EQ{Hermite} is invariant under permutations of its indices. In~\cite{Grad1949}, uniqueness of the coefficients in~\EQ{GradDist} is restored by imposing auxiliary symmetry conditions on the coefficients.

Grad's moment systems can be conveniently conceived of as Galerkin approximations of the Boltzmann equation in renormalized form in accordance with~\EQ{GalLevMomCls}. 
For a prescribed Maxwellian, the renormalization map simply corresponds to $\beta:{}g\mapsto\MM{}g$. Incorporation of the Maxwellian in~\EQ{GradDist} in the approximation can be represented by the renormalization map:
\begin{equation}
\label{eq:betaG}
\beta:g\mapsto\exp(\Pi_{\IE}g)\times\big(1+(\mathrm{Id}-\Pi_{\IE})g\big)
\end{equation}
where $\Pi_{\IE}:\IM\mapsto\IE$ denotes the orthogonal projection onto the space of collision invariants and $\mathrm{Id}$ represents the identity operator. The
embedding $\IE\subseteq\IM$ implies that $\Pi_{\IE}\IM=\IE$ and $(\mathrm{Id}-\Pi_{\IE})\IM=\IM\setminus\IE$. Hence, the projection in~\EQ{betaG} provides a
separation of~$\IM$ into~$\IE$ and its orthogonal complement. It is notable that the renormalization map in Grad's moment system
can be conceived of as a linearization of Levermore's exponential closure relation in the vicinity of $\MM$. In particular, setting $\psi=\log\MM\in\IE$,
the following identities hold pointwise:
\begin{equation}
\label{eq:betaGexp}
\exp(g)=\exp(\psi)\exp(g-\psi)=\MM\exp(g-\psi)=\MM\big(1+(g-\psi)+O(|g-\psi|^2)\big)
\end{equation}
as $(g-\psi)\to{}0$. To derive the renormalization map $\beta:g\mapsto\MM{}g$ for prescribed Maxwellians, it suffices to note that $1+g-\psi\in\IM$. To infer the
renormalization map~\EQ{betaG} if $\MM$ is retained in the approximation, we note that setting $\psi=\Pi_{\IE}g$ and omitting the remainder in~\EQ{betaGexp} 
yields~\EQ{betaG}.  

For a prescribed (global or local) Maxwellian~$\MM$, Grad's moment systems dissipate the entropy
$\HH_{\chi^2}(f) := \frac{1}{2}\MM(f/\MM-1)^2$,
provided that $\eta_{\chi^2}$ represents an entropy density for the collision operator under consideration.
It can for example be shown that~$\eta_{\chi^2}$ is generally a suitable entropy density for collision operators 
linearized about~$\MM$ (see~\cite{Grad1965}) and for BGK collision operators. Dissipation of the entropy~$\langle\eta_{\chi^2}\rangle$ can be directly
inferred from the Galerkin formulation~\EQ{GalLevMomCls}, by noting that for $\beta(g)=\MM{}g$ it holds that:
\begin{equation}
\eta_{\chi^2}'(\beta(g))=\beta(g)/\MM-1=g-1\in\IM
\end{equation} 
Hence, $\smash[b]{\eta_{\chi^2}'}$ resides in the test space~$\IM$ in~\EQ{GalLevMomCls} and dissipation of~$\smash[b]{\langle\eta_{\chi^2}\rangle}$ follows
from Galerkin orthogonality. The entropy $\langle\eta_{\chi^2}(f)\rangle$ can be associated with the $\varphi_{\chi^2}$-divergence of~$f$ relative to~$\MM$
with $\varphi_{\chi^2}(s)=\frac{1}{2}(s-1)^2$; cf.~\EQ{PhiDiv}. Grad's moment-closure relation can in fact be obtained by minimization of the 
$\varphi_{\chi^2}$-divergence subject to the moment constraints:
\begin{equation}
\label{eq:FGrad}
\ff_{\mathrm{G}}(\bm\rho)=\argmin_{f\in\IF}\big\{\big\langle\MM\varphi_{\chi^2}(f/\MM)\big\rangle:\langle\bm m f\rangle=\bm\rho\big\}
\end{equation}
The minimization problem~\EQ{FGrad} is not impaired by 
the realizability problem inherent to~\EQ{EntMin}, because 
the moment functionals $\langle{}m(\cdot)\rangle$ are continuous
in the topology corresponding to~$\langle{}\eta_{\chi^2}\rangle$.

If the Maxwellian is retained in the approximation, then an entropy for the corresponding moment systems can be non-existent or its derivation is intractable. However, for any entropy density~$\eta$ for the collision operator, the following identity holds
by virtue of the Galerkin-orthogonality property of $\beta:=\beta(g)$ in~\EQ{GalLevMomCls}:
\begin{equation}
\label{eq:GradEnt}
\begin{aligned}
\partial_t\langle\eta(\beta)\rangle
+
\partial_{x_i}\langle{}v_i\eta(\beta)\rangle&=
\langle(\eta'(\beta)-m)\partial_t\beta\rangle
+
\langle(\eta'(\beta)-m)v_i\partial_{x_i}\beta\rangle
\\
&\phantom{=}-
\langle(\eta'(\beta)-m)\CC(\beta)\rangle
+
\langle\eta'(\beta)\CC(\beta)\rangle
\end{aligned}
\end{equation}
for arbitrary $m\in\IM$. Equation~\EQ{GradEnt} implies that solutions to Grad's moment systems dissipate any entropy~$\langle\eta\rangle$ 
for the collision operator up to $\inf_{m\in\IM}\|\eta'(\beta(g))-m\|$, in some suitable norm~$\|\cdot\|$. For example, introducing the
condensed notation $g_0=\Pi_{\IE}g$, $g_1=(\Id-\Pi_{\IE})g$ and the convex functional $\eta:\IE\times\IM\setminus\IE\to\IR$ according to
\begin{equation}
\label{eq:etag0g1}
\eta(g_0,g_1)=(g_0-1)e^{g_0}(1+g_1)+e^{g_0}g_1(1+g_1)
\end{equation}
the renormalization map in~\EQ{betaG} corresponds to $\beta(g)=e^{g_0}(1+g_1)$ and it holds that
\begin{equation}
\begin{aligned}
d\eta(g_0,g_1)&=(g_0+g_1)e^{g_0}(1+g_1)\,dg_0+(g_0+2g_1)e^{g_0}\,dg_1
\\
&=(g_0+g_1)(\partial_{g_0}\beta\,dg_0+\partial_{g_1}\beta\,dg_1)+g_1e^{g_0}\,dg_1
\\
&=(g_0+g_1)\,d\beta+g_1e^{g_0}\,dg_1
\end{aligned}
\end{equation}
Considering that $g_0+g_1\in\IM$, it follows from~\EQ{GradEnt} that if $\eta$ in~\EQ{etag0g1} is an entropy density for the collision operator, then Grad's moment systems with $\beta(\cdot)$ according to~\EQ{betaG} dissipates~$\eta$ up to $O(g_1)$ as $g_1$ vanishes (in some appropriate norm). Note that~$g_1$ vanishes at equilibrium.

Grad's moment-closure relation exhibits several fundamental deficiencies that may cause breakdown of the physical and mathematical structure of the corresponding moment-closure system for large deviations from equilibrium. First, the expansion~(\ref{eq:GradDist}) admits inadmissible, locally negative distributions.
Second, the moment systems are generally non-symmetric and hyperbolicity is not guaranteed.
It has been observed in~\cite{Brini2001,Torrilhon2000} that Grad's moment-closure systems can indeed
exhibit complex characteristics and loss of hyperbolicity. 

\section{Divergence-Based Moment Closures}
\label{sec:DivMomCls}
In this section we present a novel moment-closure relation based on an approximation of the exponential function. The considered approximation is derived from truncations of the standard limit definition of the exponential $\exp(\cdot)\defeq\smash[t]{\lim_{n\rightarrow\infty} (1 + (\cdot)/n)^n\approx(1 + (\cdot)/N)^N}$. It is noteworthy that unlike the exponential function, in the limit as $v\rightarrow-\infty$ the truncated exponential as well as its derivative do not vanish. The former condition is needed to preserve the decay properties of the exponential function while the latter condition is needed to preserve the same absolute maximum and minimum as the exponential. Moreover, as opposed to the
exponential function, the truncated exponential can be negative if $N$ is odd. 
Several approximations of the exponential function that preserve the aforementioned 
properties of the exponential have been proposed in the literature; see, for example, \cite{Tsallis2009,Naudts2011,Kaniadakis2013} and references therein. 
These so-called {\em deformed exponentials\/} can generally serve to construct  
moment-closure renormalization maps, with properties depending on the particular form of the
deformed exponential and the construction. In \cite{Tsallis2009}, Tsallis proposed the $q$-exponential:
\begin{equation}
\label{eq:ApproxExp}
\widetilde{\exp}_q (x) \defeq \big(1+ (1-q)x\big)^{1/(1-q)}_+
\end{equation}
with $q\neq{}1$ and $(\cdot)_+ = \frac{1}{2}(\cdot)+\frac{1}{2}|\cdot|$ the non-negative part of a function extended by~$0$. The
$q$\nobreakdash-exponential in~\EQ{ApproxExp} is related to the non-negative part of the truncated limit definition of the
exponential by $1-q=1/N$. We will consider renormalization maps of the form
\begin{equation}
\label{eq:BTsallis}
\beta_N:g\mapsto\mathcal{M}\,\widetilde{\exp}_q(g)=\mathcal{M}\Big(1+\frac{g}{N}\Big)^N_+
\end{equation}
with $\mathcal{M}$ a prescribed local or global Maxwell distribution. The renormalization map $\beta_N$ can be construed as an approximation
to the exponential renormalization map about the Maxwellian distribution~$\mathcal{M}$. 
We will establish that the moment-closure distribution (\ref{eq:BTsallis}) can be derived as the 
minimizer of a modified entropy that approximates the Kullback--Leibler divergence near~$\mathcal{M}$ and that belongs to the class 
of $\varphi$\nobreakdash-divergences. In
addition, we will show that the resulting moment system overcomes the aforementioned deficiencies of Grad's and Levermore's moment systems, while retaining the fundamental properties of the Boltzmann equation presented in Section~\ref{sec:BoltzProps}.

The renormalization map~\EQ{BTsallis} engenders the following moment-closure relation:
\begin{equation}
\label{eq:DivCls}
\mathcal{F}_N(\bm\rho)\defeq 
 \mathcal{M} \ \widetilde{\exp}_q(\bm\alpha\cdot \bm m)
 =
\mathcal{M}
\left(1+\frac{\bm\alpha\cdot \bm m}{N}\right)^N_+
\end{equation}
where the moment densities $\bm{\rho}$ and the coefficients $\bm\alpha$ are related by $\bm\rho=\langle\bm{m}\,\mathcal{M}\,\widetilde{\exp}_q(\bm\alpha\cdot\bm{m})\rangle$. Given a polynomial subspace $\mathbb{M}\supseteq\IE$ with a Galilean-group property (admissibility conditions 1 and 2 in section \ref{sec:LevMomCls}),
the moment system corresponding to~\EQ{BTsallis} conforms to~\EQ{ClsMomSys} with, in particular,
the moment-closure relation $\mathcal{F}_N$ according to~(\ref{eq:DivCls}).

To elucidate some of the characteristics of the renormalization map~\EQ{BTsallis}, we regard it in comparison with the renormalization maps associated
with  Levermore's exponential moment-closure relation and Grad's moment-closure relation with a prescribed Maxwellian prefactor. 
The renormalization map associated with Levermore's moment-closure relation is given by $g\mapsto\exp(g)$; see Section~\SEC{LevMomCls}. 
By virtue of the vector-space structure of $\IM\supseteq\IE$, for an arbitrary Maxwellian distribution~$\mathcal{M}$ it holds that~$\log\mathcal{M}+\IM=\IM$. Hence, for 
$g\in\IM$, the renormalization map $g\mapsto\exp(g)$ can be equivalently expressed as $g\mapsto\exp(\log\mathcal{M}+g)$.
In the limit $N\to\infty$, we obtain for~\EQ{BTsallis}:
\begin{equation}
\label{eq:BTsallisLim}
\lim_{N\to\infty}\beta_N(g)=\mathcal{M}\lim_{N\to\infty}\Big(1+\frac{g}{N}\Big)^N_+=\mathcal{M}\exp(g)=\exp(\log\mathcal{M}+g)
\end{equation}
Equation~\EQ{BTsallisLim} implies that in the limit $N\to\infty$, the renormalization map in~\EQ{BTsallis} coincides with the exponential renormalization map
associated with Levermore's moment-closure relation.
For finite~$N$, the moments $\langle{}m\beta_N(g)\rangle$ and fluxes $\langle{}mv\beta_N(g)\rangle$ with $m,g\in\IM$ correspond to piecewise-polynomial 
moments of the Gaussian distribution~$\mathcal{M}$. The evaluation of such moments is tractable, as opposed to the evaluation of moments and fluxes 
for the exponential renormalization map.  In addition, for super-quadratic approximations $\IM\supset\IE$, the exponential renormalization map associated 
with Levermore's closure can lead to singular moments and fluxes in the vicinity of equilibrium, i.e. as~$g$ approaches $\IE$. The fundamental
underlying problem is the realizability problem; see Section~\SEC{LevMomCls} and~\cite{Junk1998}. Accordingly, one can form sequences $\{g_n\}$ 
such that $\exp(g_n)\to\IE$ (in the $L^1$~topology) while there exist $m\in\IM$ such 
that~$|\langle{}m{}\exp(g_n)\rangle|\to\infty$ or~$|\langle{}m{}v\exp(g_n)\rangle|\to\infty$. One can infer that due to the exponential decay of the prefactor $\mathcal{M}$ 
and the polynomial form of the renormalization map in~\EQ{BTsallis}, moments and fluxes corresponding to~\EQ{BTsallis} are non-singular near equilibrium.
To compare~\EQ{BTsallis} to the renormalization map
corresponding to Grad's moment-closure relation with a prescribed Maxwellian prefactor, $g\mapsto\mathcal{M}g$ (see Section~\SEC{Grad}), 
we note that by virtue of the vector-space structure of~$\IM\supseteq\IE$, it holds that $1+\IM=\IM$. Hence, for $g\in\IM$, the renormalization map $g\mapsto\mathcal{M}g$ 
can be equivalently expressed as $\beta_{\mathrm{G}}:g\mapsto\mathcal{M}(1+g)$. Comparison of $\beta_N$ and $\beta_{\mathrm{G}}$  imparts that
$\beta_1=(\beta_{\mathrm{G}})_+$, i.e. for $N=1$ the renormalization map~$\beta_N$ in~\EQ{BTsallis} coincides with the non-negative part of the renormalization
in Grad's closure, extended by zero. Therefore, the renormalization map~\EQ{BTsallis} avoids the potential negativity of the approximate
distribution inherent to Grad's closure and the corresponding loss of hyperbolicity of the moment system. 

The moment system corresponding to~\EQ{BTsallis} retains conservation of mass, momentum and energy as well as Galilean invariance. 
The conservation properties can be directly deduced from the Galerkin form~\EQ{GalLevMomCls} of the moment system, by noting that~$\IE$ is contained in
the test space~$\IM$, in accordance with admissibility condition 1 in section~\ref{sec:LevMomCls}). Galilean invariance is an immediate consequence of admissibility condition~2. However, contrary to Levermore's moment system, the moment system with renormalization map~\EQ{BTsallis} does not generally dissipate the 
relative entropy $\langle{}f\log{}(f/\mathcal{M})\rangle$, because the inverse of $\beta_N(\cdot)$ does not correspond to $\log(\cdot)$ and, therefore, 
$\log\beta_N(g)$ does not generally belong to the test space~$\IM$ for~$g\in\IM$; cf. Section~\SEC{LevMomCls}.
The moment system closed by~\EQ{BTsallis} does however dissipate a modified entropy. To determine a suitable entropy function for the moment system with renormalization map~\EQ{BTsallis}, we observe that:
\begin{equation}
\label{eq:approxinv}
\beta_N^{-1}(\cdot)=N\bigg(\frac{(\cdot)}{\mathcal{M}}\bigg)^{1/N}-N=:\widetilde{\log}_N\big((\cdot)/\mathcal{M}\big)
\end{equation}
provides an inverse of $\beta_N$ according to~\EQ{BTsallis} with domain $\IR_{\geq0}$. The function $\smash[t]{\widetilde{\log}}$ yields an approximation to the 
natural logarithm, corresponding to the inverse of the $q$\nobreakdash-exponential in~\EQ{ApproxExp}. The approximate logarithm is eligible as the derivative
of an entropy density associated with the moment system with renormalization~\EQ{BTsallis}. In particular, defining the entropy density as
\begin{equation}
\label{eq:approxeta}
\eta_N(f)=f\bigg(\frac{N^2}{1+N}\bigg(\frac{f}{\mathcal{M}}\bigg)^{1/N} - N\bigg)+\mathcal{M}\frac{N}{1+N}
\end{equation}
it holds that $\eta_N'=\beta_N^{-1}$ and, hence, $\eta_N'(\beta_N(\cdot)):\IM\to\IM$. The constant in~\EQ{approxeta} has been selected
such that $\eta_N(\mathcal{M})$ vanishes. The entropy corresponding to~\EQ{approxeta} can be cast in the form of a relative entropy associated with
a $\varphi$\nobreakdash-divergence, in accordance with~\EQ{PhiDiv}. To this end, we introduce
\begin{equation}
\label{eq:varphiN}
\varphi_N(\cdot)=(\cdot)\bigg(\frac{N^2}{1+N}(\cdot)^{1/N} - N\bigg)+\frac{N}{1+N}
\end{equation}
and note that $\eta_N(f)=\mathcal{M}\varphi_N(f/\mathcal{M})$. Convexity of the function~$\varphi_N$ and of the corresponding entropy density~$\eta_L$ follows by 
direct computation:
\begin{equation}
\varphi_N''(\cdot)=(\cdot)^{1/N}-1
\end{equation}
Therefore, $\varphi_N''$ is strictly positive on $\IR_{>0}$. Moreover, it holds that $\varphi_N(1)=0$. In conclusion, if 
$\eta_N(\cdot)=\mathcal{M}\varphi((\cdot)/\mathcal{M})$ 
is an entropy density for the collision operator~$\CC$ according to~\EQ{Dissipation}, then the approximate distribution~$\beta_N(g)$
of the moment system~\EQ{GalLevMomCls} with renormalization map~\EQ{BTsallis} complies with the local entropy-dissipation relation:
\begin{equation}
\partial_t\big\langle\eta_N\big(\beta_N(g)\big)\big\rangle
+
\partial_{x_i}\big\langle{}v_i\eta_N\big(\beta_N(g)\big)\big\rangle
=
\big\langle\CC(\beta_N(g))\eta_N'\big(\beta_N(g)\big)\big\rangle
\leq
0
\end{equation}
We recall that the premise on the collision operator is for example satisfied by the BGK and extended BGK operators; see also Remark~\ref{rem:remark1}.

The moment-closure relation~\EQ{DivCls} can be derived by minimization of the $\varphi_N$-divergence subject to the moment constraint; cf. the definition
of Levermore's closure relation according to~\EQ{EntMin}. Consider the constrained minimization problem:
\begin{equation}
\label{eq:AppEntMin}
\FN(\bm\rho):=
\argmin_{f\in\mathbb{F}}
\big\{\langle \eta_N(f)\rangle:\langle \bm m f \rangle=\bm\rho\big\}
\end{equation}
Formally, the solution to~\EQ{AppEntMin} can be obtained by the method of Lagrange multipliers.  The minimizer in~\EQ{AppEntMin}
corresponds to a stationary point of the Lagrangian $(f,\bm\alpha)\mapsto\langle\eta_N(f)\rangle+\bm\alpha\cdot(\bm\rho-\langle{\bm m}f\rangle)$.
The stationarity condition implies that $\eta_N'(f)-\bm\alpha\cdot\bm m=0$ and, on account of~\EQ{approxinv}, that $\beta_N^{-1}(f)=\bm\alpha\cdot\bm m$.
It follows directly that the minimizer in~\EQ{AppEntMin} is of the form $\FN(\bm\rho)=\beta_N(\bm\alpha\cdot\bm m)$ in conformity with~\EQ{DivCls}.
Contrary to the entropy minimization problem~\EQ{EntMin} underlying Levermore's closure relation, the minimization problem~\EQ{AppEntMin}
is well posed. Existence of a solution to the minimization problem~\EQ{AppEntMin} can be deduced from results for generalized projections
for non-negative functions by Csisz\'ar in~\cite{Csiszar1995}. In \cite{Csiszar1995} it is shown that the minimization problem
\begin{equation}
\label{eq:Csiszar1}
\inf\bigg\{\int{}f_2(v)\,\varphi\big(f_1(v)/f_2(v)\big)\,\nu(d{}v):f_1\in\IX\bigg\}
\end{equation}
over a constrained set of non-negative functions,
\begin{equation}
\label{eq:IX}
\IX=\bigg\{f:\int{}a_j(v)\,f(v)\,\nu(d{}v)=\rho_j,\,j\in\IJ\bigg\}
\end{equation}
for certain countable functions $\{a_j\}_{j\in\IJ}$, possesses a minimizer belonging to~$\IX$ provided that the following (sufficient) conditions hold:
\begin{enumerate}
\renewcommand{\theenumi}{\arabic{enumi}}
\renewcommand{\labelenumi}{\theenumi)}
\item $\IX$ is a convex set of non-negative functions and the infimum in~\EQ{Csiszar1} is finite;
\label{itm:cond1}
\item $\lim_{s\rightarrow\infty} \varphi'(s) = \infty$;
\label{itm:cond2}
\item $\int\varphi^{\star}(\xi|a_j(v)|)f_2(v)\,\nu(dv)$ is finite for all~$\xi>0$ and $j\in\IJ$.
\label{itm:cond3}
\end{enumerate}
The function $\varphi^{\star}$ in condition~\ref{itm:cond3} corresponds to the convex conjugate of~$\varphi$. Comparison conveys that~\EQ{AppEntMin} 
conforms to~\EQ{Csiszar1}\nobreakdash--\EQ{IX} with $f_2=\MM$, $\nu(\cdot)$ Lebesgue measure and $\{a_j\}_{j\in\IJ}$ a monomial basis of~$\IM$.
Convexity of the constrained distributions follows from the linearity of the moment constraints. Finiteness of the infimum is ensured by the fact that
the infimum over the constrained set is bounded from below by the infimum over the unconstrained set, and the latter attains its minimum of $0$ for~$f=\MM$.
The minimization problem~\EQ{AppEntMin} thus complies with condition~\ref{itm:cond1}.
Compliance with condition~\ref{itm:cond2} follows from $\varphi_N'(s)=Ns^{1/N}-N$ and $\lim_{s\to\infty}s^{1/N}=\infty$. To verify condition~\ref{itm:cond3}, we note that
the convex conjugate of~$\varphi_N$ is:
\begin{equation}
\varphi_N^{\star}(t)=\sup_{s\in\IR_{\geq0}}\big(st-\varphi_N(s)\big)=\frac{N}{1+N}\bigg(\bigg(1+\frac{t}{N}\bigg)^{N+1}_+-1\bigg)
\end{equation}
Condition~\ref{itm:cond2} therefore translates into the requirement that
\begin{equation}
\label{eq:Cond3ver}
\big\langle\MM\varphi^{\star}_N\big(\xi|m_j|\big)\big\rangle
=
\frac{N}{1+N}\bigg\langle\MM\bigg(1+\frac{\xi|m_j|}{N}\bigg)^{N+1}_+-\MM\bigg\rangle
\end{equation}
is bounded. By virtue of the exponential decay of the prefactor~$\MM$ and the fact that $|m_j|^{N+1}$ increases only polynomially, the expressions
in~\EQ{Cond3ver} are indeed finite for any $\xi>0$. The minimization problem~\EQ{AppEntMin} therefore also satisfies condition~\ref{itm:cond3}. It is
notable that the minimization problem~\EQ{EntMin} pertaining to Levermore's moment closure satisfies conditions~\ref{itm:cond1} and~\ref{itm:cond2}
but violates condition~\ref{itm:cond3}.

To establish that the closure relation~\EQ{DivCls} leads to a symmetric-hyperbolic system, we insert~\EQ{DivCls} into the
generic form~\EQ{ClsMomSys} of moment systems, and note that application of the chain rule and product rule leads to a system of the
form~\EQ{HypSys} with:
\begin{subequations}
\label{eq:SymmHyp}
\begin{align}
{\bm A}_0(\bm\alpha)
&=\left\langle\bm m \otimes\bm m\, \mathcal{M}\left(1+\frac{\bm\alpha\cdot\bm m}{N}\right)_+^{N-1}\right\rangle 
\\
{\bm A}_i(\bm\alpha)
&=
\left\langle v_i\bm m \otimes \bm m \,\mathcal{M}\left(1+\frac{\bm\alpha\cdot\bm m}{N}\right)_+^{N-1}\right\rangle 
\\
{\bm s}(\bm \alpha)
&= 
\bigg\langle\bm m \,
\mathcal{C}\bigg(\mathcal{M}\left(1+\frac{\bm\alpha\cdot\bm m}{N}\right)_+^{N}\bigg)\bigg\rangle 
-
\bigg\langle\bm m \, \left(1+\frac{\bm\alpha\cdot\bm m}{N}\right)_+^{N} \bigg(\frac{\partial\MM}{\partial{}t}+\sum_{i=1}^Dv_i\frac{\partial\MM}{\partial{}x_i}\bigg)\bigg\rangle
\label{eq:sourceterm}
\end{align}
\end{subequations}
The symmetry of ${\bm A}_0,\ldots,{\bm A}_D$ is evident. Positive-definiteness of ${\bm A}_0$ follows from:
\begin{equation}
\label{eq:SPDprop}
\bm\gamma\cdot\left\langle\bm m \otimes \bm m \,\mathcal{M}\left(1+\frac{\bm\alpha\cdot\bm m}{N}\right)^{N-1}_+\right\rangle\bm\gamma 
=
\left\langle(\bm\gamma\cdot\bm m )^2\mathcal{M}\left(1+\frac{\bm\alpha\cdot\bm m}{N}\right)^{N-1}_+\right\rangle\geq{}0
\end{equation}
The inequality in~\EQ{SPDprop} reduces to an equality if and only if $\bm\gamma=0$ or $\bm\alpha\cdot\bm m =-N$. The latter case is 
pathological, because  $\bm\alpha\cdot\bm m =-N$ implies that $\FN(\bm\rho)=0$. 

It is noteworthy that the second constituent of the production term~${\bm s}({\bm\alpha})$, i.e. the term representing the contribution
of $\partial_t\MM+v_i\partial_{x_i}\MM$ to the production, may cause blow up of solutions to the hyperbolic system~\EQ{ClsMomSys} with~\EQ{SymmHyp}
in the limit $t\rightarrow\infty$. Hence, the hyperbolic character of~\EQ{ClsMomSys} with~\EQ{SymmHyp} ensures stability of solutions only in finite time. 
If $\mathcal{M}$ corresponds to a global Maxwellian, then  $\partial_t\MM+v_i\partial_{x_i}\MM$ vanishes and the stability provided by hyperbolicity also
holds in the ad-infinitum limit.

\section{Numerical results for the 1D spatially homogeneous Boltzmann-BGK equation}
\label{sec:NumExp}
To illustrate the properties of the moment-system approximation~\EQ{GalLevMomCls} with the divergence-based closure relation
encoded by the renormalization map~\EQ{BTsallis}, this section presents numerical computations for the spatially homogeneous Boltzmann-BGK equation in~1D:
\begin{subequations}
\label{eq:BGKapprox}
\begin{align}
\partial_tf &=-\tau^{-1}(f-\mathcal{E}_f)
\label{eq:BGKODE}
\\
f(0,v)&=f_0(v)
\end{align}
\end{subequations}
for some given initial distribution~$f_0$. The corresponding moment system writes:
\begin{subequations} 
\label{eq:BGKMomApprox}
\begin{align}
\partial_t \langle \bm m \FN \rangle &= -\tau^{-1}\big\langle \bm m( \mathcal{F}_N-\mathcal{E}_{ \mathcal{F}_N})\big\rangle
\label{eq:MomODE}
\\ 
\FN(0,v) &= (\FN)_0(v) 
\end{align}
\end{subequations}
with $\FN$ according to~\EQ{DivCls} and $(\FN)_0$  defined by the minimization problem~\EQ{AppEntMin} subject to the moments
corresponding to the initial distribution:
\begin{equation}
\label{eq:FN0}
\begin{aligned}
(\FN)_0 \defeq&\argmin_{f\in\IF}\big\{\langle\eta_N(f)\rangle:\langle \bm m f \rangle=\langle\bm m  f_0 \rangle\big\}
\\
=&\big\{f\in\IF:f=\beta_N(g),g\in\IM,\langle \bm m f \rangle=\langle\bm m  f_0 \rangle\big\}
\end{aligned}
\end{equation}
The systems in \EQ{BGKapprox} and \EQ{BGKMomApprox} represent initial-value problems for 
the ordinary differential equations~\EQ{BGKODE} and~\EQ{MomODE}. The solutions of 
the initial-value problems \EQ{BGKapprox} and \EQ{BGKMomApprox} are, respectively,
\begin{subequations}
\label{eq:HomBGKsol}
\begin{align}
f(t,v) &=e^{-(t-t_0)/\tau} f_0 + \big(1-e^{-(t-t_0)/\tau}\big)\mathcal{E}_f 
\\
\mathcal{F}_N(t,v) &=  e^{-(t-t_0)/\tau} (\FN)_0 + \big(1-e^{-(t-t_0)/\tau}\big) \mathcal{E}_{\mathcal{F}_N}
\end{align}
\end{subequations}
For the considered spatially-homogeneous case, the collision-invariance properties of the collision operator imply that $\mathcal{E}_{f}=\smash[b]{\mathcal{E}_{f_0}}$ 
and, similarly, $\smash[b]{\mathcal{E}_{\FN}=\mathcal{E}_{(\FN)_0}}$. Furthermore, the constraints in the minimization problem in~\EQ{FN0} impose
$\smash[b]{\mathcal{E}_{f_0}= \mathcal{E}_{(\FN)_0}}$. Based on the expressions for the solutions in~\EQ{HomBGKsol}, it then follows that:
\begin{equation}
\label{eq:normdiff}
\big\|f(t,\cdot)-\FN(t,\cdot)\big\| = e^{-(t-t_0)/\tau}\big\|f_0 -(\FN)_0\big\|
\end{equation}
in any suitable norm.
Equation~\EQ{normdiff} conveys that the accuracy of the approximation of $f(t,\cdot)$ by $\FN(t,\cdot)$ at any time $t>0$ depends exclusively on the accuracy 
of the approximation of the initial condition $f_0$ by~$(\FN)_0$ according to~\EQ{FN0}. In the remainder of this section we therefore restrict our considerations to 
numerical examples that illustrate the approximation properties of $\varphi_N$\nobreakdash-divergence minimizers and to properties of the projection problem~\EQ{FN0}.

\begin{figure}
\begin{center}
\includegraphics[width=1\textwidth]{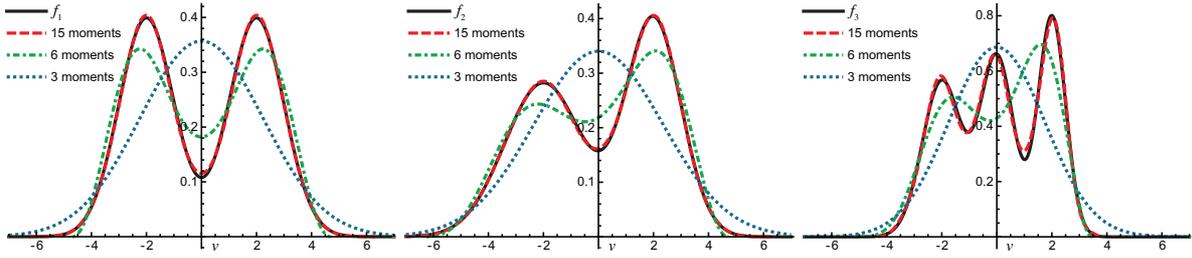}
\end{center}
\caption{Distributions $f_1$ ({\em left\/}), $f_2$ ({\em center\/}) and $f_3$ ({\em right\/}) according to~\EQ{sampledist} and the corresponding approximations 
with $k=3$ ({\em dotted\/}), $k=6$ ({\em dash-dot\/}) and $k=15$ ({\em dashed\/}) moments obtained from the moment-constrained $\varphi$-divergence minimization problem~\EQ{FN0}.
\label{fig:initDist}}
\end{figure}

\begin{figure}
\begin{center}
\includegraphics[width=0.85\textwidth]{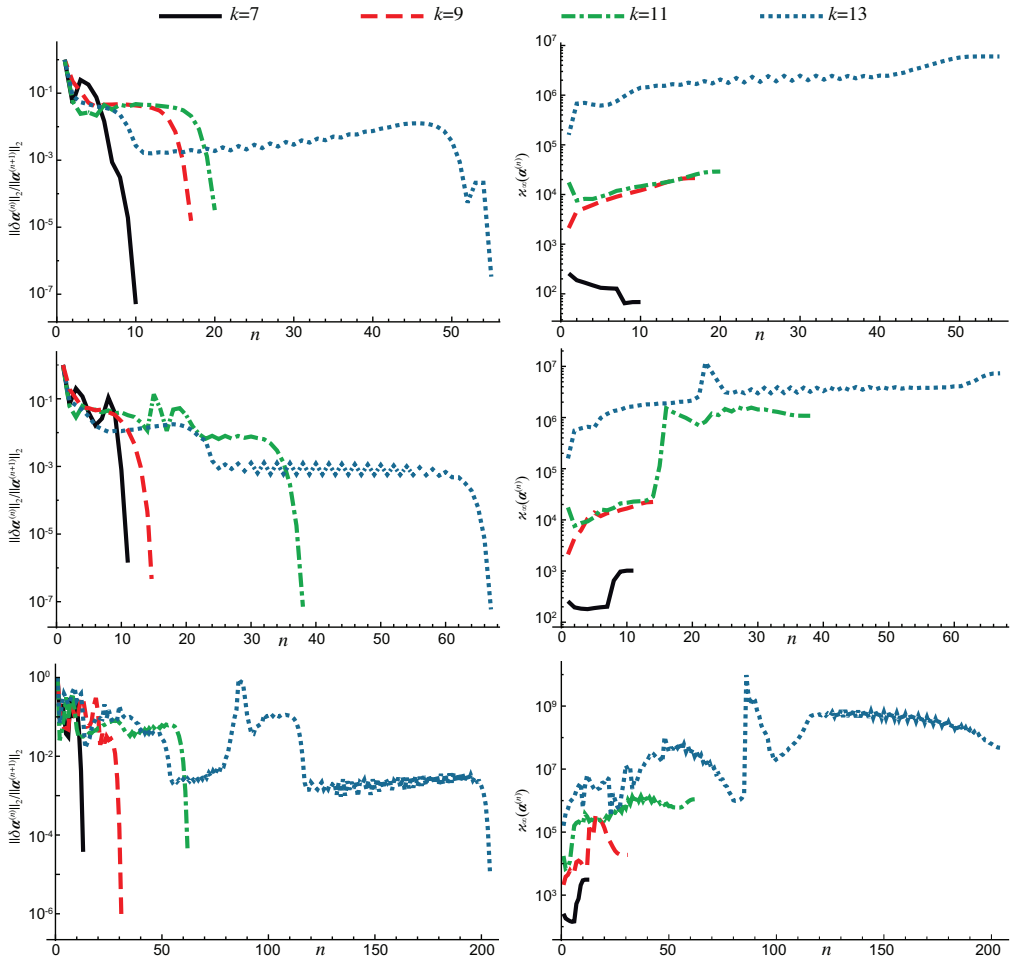}
\end{center}
\caption{%
Convergence of the Newton process for the nonlinear projection problem~\EQ{NonlinProj} and conditioning of the corresponding Jacobian matrices:
({\em left\/}) relative magnitude of the Newton update, $\|\delta\bm\alpha\|_2/\|\bm\alpha^{(n+1)}\|_2$, versus the number of iterations
for $k=7,9,11,13$ and for distributions~\EQ{sampledist1} ({\em top\/}), \EQ{sampledist2} ({\em center\/}) and~\EQ{sampledist3} ({\em bottom\/});
({\em right\/}) corresponding $\infty$\nobreakdash-norm condition numbers, $\varkappa_{\infty}({\bm\alpha}^{(n)})$, of the Jacobian matrices according to~\EQ{Jacobian}.
\label{fig:Newton}}
\end{figure}

We consider approximations of the distributions: 
\begin{subequations}
\label{eq:sampledist}
\begin{align}
f_1(v) &= \frac{e^{-\frac{1}{2}(v - 2)^2}}{\sqrt{2 \pi}} +\frac{e^{-\frac{1}{2}(v + 2)^2}}{\sqrt{2 \pi}},
\label{eq:sampledist1}\\
f_2(v) &= \frac{e^{-\frac{1}{2}(v - 2)^2}}{\sqrt{2 \pi}} +\frac{e^{-\frac{1}{4}(v + 2)^2}}{\sqrt{4 \pi}},
\label{eq:sampledist2}\\
f_3(v) &= \frac{e^{-2(v - 2)^2}}{\sqrt{\pi/2}} +\frac{e^{-\frac{4}{3}v^2}}{\sqrt{3\pi/4}}+\frac{e^{-(v + 2)^2}}{\sqrt{\pi}}
\label{eq:sampledist3}
\end{align}
\end{subequations}
by means of moment-constrained $\varphi_N$\nobreakdash-divergence 
minimizers in polynomial spaces of increasing order. The distributions in~\EQ{sampledist1}--\EQ{sampledist3}, shown in Figure \ref{fig:initDist}, correspond to 
distributions of increasing complexity, viz.,
a symmetric bi-modal distribution, a non-symmetric bi-modal distribution and a 
non-symmetric tri-modal distribution, respectively. 
For the pre-factor $\MM$ in the renormalization map~\EQ{BTsallis} and, accordingly, in the
relative entropy $\langle\eta_N(f)\rangle=\langle\MM\varphi_N(f/\mathcal{M})\rangle$
associated with the $\varphi_N$\nobreakdash-divergence in~\EQ{varphiN},
we select the global equilibrium distribution~$\smash[b]{\mathcal{E}_{f_0}}$.
In particular, denoting by $\IM_k=\mathrm{span}\{1,v,\ldots,v^{k-1}\}$ the space of polynomials of degree $k-1$, the minimization problem in~\EQ{FN0}
with $k$~moment ($k\geq{}3$) constraints engenders the nonlinear-projection problem:
\begin{equation}
\label{eq:NonlinProj}
(\FN)_0=\MM\left(1+\frac{g}{N}\right)^{N}_+,\,g\in\IM_k:
\qquad
\bigg\langle{}m\MM\Big(1+\frac{g}{N}\Big)^{N}_+\bigg\rangle
=
\big\langle{}mf_0\big\rangle
\qquad
\forall{}m\in\IM_k
\end{equation}
Expanding $g(v)=\alpha_i{}v^{i-1}=\bm\alpha\cdot{\bm m}(v)$, Equation~\EQ{NonlinProj} corresponds to a nonlinear algebraic system for the
coefficients $\bm\alpha$. To evaluate the integrals in \EQ{BGKMomApprox}, we first determine the roots of the 
polynomial $(1+\bm\alpha\cdot{\bm m}/N)$ and then establish the limits of the positive parts to compute the corresponding contributions to the moments of $\MM(1+\bm\alpha\cdot{\bm m}/N)_+^N$. The integrals are evaluated by applying a suitable transformation of the integration variable and invoking the following rule:
\begin{equation} 
\label{eq:intRule}
\begin{aligned}
\int_{v_0}^{v_1} e^{-v^2} v^k \, dv &= \frac{1}{2} \left( \Gamma\left(\frac{1+k}{2}\right)-\Gamma\left(\frac{1+k}{2},v_1^2\right)\right)\text{sign}^{1+k}(v_1)
\\
&\phantom{=}
-\frac{1}{2}\left( \Gamma\left(\frac{1+k}{2}\right)-\Gamma\left(\frac{1+k}{2},v_0^2\right)\right)\text{sign}^{1+k}(v_0)
\end{aligned}
\end{equation}
where $-\infty\leq v_0 \leq \infty$ and $-\infty\leq v_1 \leq \infty$ and $\Gamma(\cdot)$ and $\Gamma(\cdot,\cdot)$ are the complete and incomplete gamma functions respectively. The coefficients~$\bm\alpha$ are extracted from the system~\EQ{NonlinProj}  by means of the
Newton method. It is to be noted that $(1+g/N)^N_+$ is Fr\'echet differentiable with respect to~$g$ by virtue of the fact that, evidently, 
changes in the sign of $1+g/N$ occur only at roots. A consistent Jacobian for the tangent problems in the Newton method is provided by:
\begin{equation}
\label{eq:Jacobian}
\frac{d}{d\bm\alpha}
\bigg\langle{\bm m}\,\MM\Big(1+\frac{\bm\alpha\cdot{\bm m}}{N}\Big)^N_+\bigg\rangle
=
\bigg\langle{\bm m}\otimes{\bm m}\,\MM\Big(1+\frac{\bm\alpha\cdot{\bm m}}{N}\Big)^{N-1}_+\bigg\rangle
=:{\bm J}({\bm\alpha}).
\end{equation}
%
The Jacobian matrix ${\bm J}({\bm\alpha})$ in the right member of~\EQ{Jacobian} can be identified as a symmetric-positive definite matrix and, hence, the tangent problems in the Newton method are well posed. The Jacobian is however of Hankel-type and it becomes increasingly ill-conditioned as the number of moments increases; 
see for example \cite{Fasino1995,Taylor1978}. Consequently, the convergence behavior of the Newton process deteriorates for higher-moment systems.
To illustrate the dependence of the convergence behavior of the Newton process on the number of moments, 
Figure~\ref{fig:Newton} ({\em left\/}) plots the ratio of the $2$\nobreakdash-norm of the update in the Newton process, $\|\delta\bm\alpha\|_2$,
over the  $2$\nobreakdash-norm of the solution vector, $\|\bm\alpha^{(n+1)}\|_2$,
versus the number of iterations for polynomial orders $k=7,9,11,13$ for the three test distributions in~\EQ{sampledist}. 
The ratio $\|\delta\bm\alpha\|_2/\|\bm\alpha^{(n+1)}\|_2$ can be conceived of as the relative magnitude of the update vector.
Figure~\ref{fig:Newton} ({\em right\/}) 
plots an approximation of the corresponding infinity-norm condition numbers, $\varkappa_{\infty}({\bm\alpha})=\|{\bm J}({\bm\alpha})\|_{\infty}\|{\bm J}^{-1}({\bm\alpha})\|_{\infty}$, of the Jacobian matrices. 
The results in Figure~\ref{fig:Newton} convey that the condition number increases significantly as the number of moments increases.
For $k=7$ the condition number is approximately $10^3$, while for $k=13$ the condition number exceeds $10^5$ and can even reach~$10^{10}$.
For high-order approximations, the convergence behavior of the Newton process is generally slow and non-monotonous. However, in all cases the 
relative update can be reduced to a tolerance of~$10^{-4}$.

\begin{figure}
\begin{center}
\includegraphics[width=\textwidth]{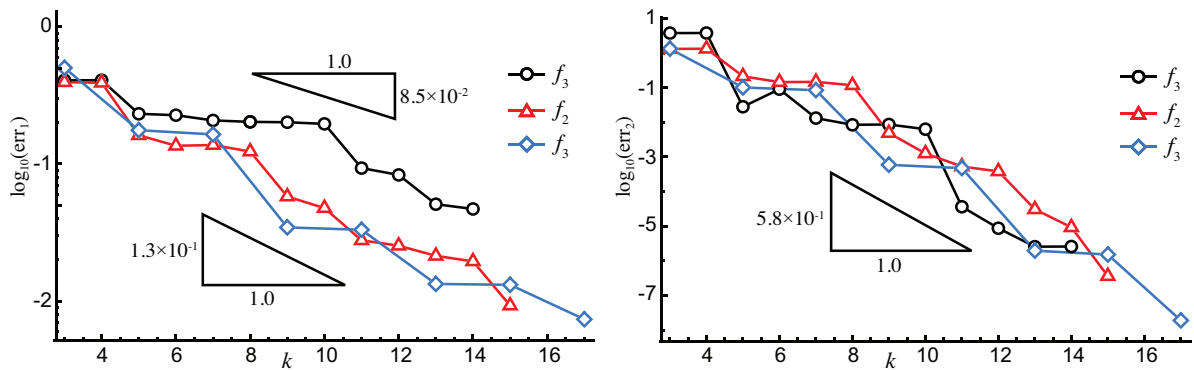}
\end{center}
\caption{%
Approximation properties of the moment method with closure relation~\EQ{DivCls}: ({\em left\/}) convergence of 
the relative error in the $L^1(\IR)$\nobreakdash-norm $\mathrm{err}_1$ according to~\EQ{errmeas} for $f_1,f_2$ and $f_3$ 
in~\EQ{sampledist}; ({\em right\/}) corresponding convergence of the relative error in the cosine moment, $\mathrm{err}_2$.
\label{fig:Convergence}}
\end{figure}

To illustrate the approximation properties of the moment method with closure relation~\EQ{DivCls}, Figure~\ref{fig:Convergence} ({\em left\/}) plots the
$L^1(\IR)$\nobreakdash-norm of the relative error in the approximation $\mathcal{F}_2^i$ to the distribution~$f_i$ ($i=1,2,3$) according to~\EQ{sampledist} 
and Figure~\ref{fig:Convergence} ({\em right\/}) the corresponding relative error in the cosine moment,
\begin{equation}
\label{eq:errmeas}
\mathrm{err}_1=
\frac{\|f_i-\mathcal{F}^i_2\|_{{L}^1(\mathbb{R})}}{\|f_i\|_{\mathbb{L}^1(\mathbb{R})}},
\qquad
\mathrm{err}_2=\frac{|\langle \cos(\cdot) f_i\rangle- \langle\cos(\cdot)\mathcal{F}^i_2\rangle|}{|\langle \cos(\cdot)f_i\rangle|},
\end{equation}
respectively. The cosine moment serves to investigate the super-convergence properties of the approximation in accordance with the
Babu\v{s}ka--Miller theorem~\cite{Babuska:1984fk}; see Section~\ref{sec:MomSys}. A non-polynomial moment has been selected
to examine the convergence behavior, because for any polynomial moment $\langle\sigma{}f_i\rangle$ with $\sigma\in\IM_l$ the approximation
$\langle\sigma\mathcal{F}_2^i\rangle$ provided by the $k$\nobreakdash-moment approximation~$\mathcal{F}^i_2$ is exact for all $k\geq{}l$,
on account of the constraints in~\EQ{FN0}. Figure~\ref{fig:Convergence} ({\em left\/}) indicates that $\|f_i-\mathcal{F}_2^i\|_{L^1(\IR)}$
converges exponentially with increasing~$k$, i.e., there exist positive constants~$C$ and~$\zeta$ such that $\|f_i-\mathcal{F}_2^i\|_{L^1(\IR)}\leq{}C\zeta^{-k}$.
In particular, $\zeta\approx{}10^{-0.13}\approx0.74$ for the bi-modal distributions~$f_1$ and~$f_2$ and $\zeta\approx10^{-0.085}\approx0.82$ for the tri-modal distribution~$f_3$.
Comparison of the left and right panels in Figure~\ref{fig:Convergence} conveys that the approximation of the cosine moment indeed
converges at a higher rate than the $L^1(\IR)$\nobreakdash-norm of the approximation itself. Figure~\ref{fig:Convergence} ({\em right\/})
 conveys that the cosine moment converges at a rate of~$\zeta\approx10^{-0.58}\approx0.26$ for both the bi-modal distributions $f_1,f_2$ and the tri-modal distribution~$f_3$. 

\section{Conclusion}
\label{sec:Conc}
To avoid the realizability problem inherent to the maximum-entropy closure relation for moment-system approximations of the Boltzmann equation, we proposed a class of new closure relations based on $\varphi$\nobreakdash-divergence minimization. We established
that $\varphi$-divergences provide a natural generalization of the usual relative-entropy setting of the moment-closure problem. 
It was shown that minimization of certain $\varphi$\nobreakdash-divergences leads to suitable closure relations and that the corresponding moment-constrained 
$\varphi$\nobreakdash-divergence minimization problems are not impaired 
by the realizability problem inherent to relative-entropy minimization. Moreover,  
if the collision 
operator under consideration dissipates a $\varphi$\nobreakdash-divergence,
then the corresponding minimal-divergence moment-closure systems retain the fundamental structural properties of 
the Boltzmann equation, namely, conservation of mass, momentum and energy, Galilean invariance, and dissipation of an entropy, sc. the $\varphi$\nobreakdash-divergence. 
For suitable $\varphi$\nobreakdash-divergences, the closure relation yields
non-negative approximations of the one-particle marginal.
Divergence-based moment systems are generally symmetric hyperbolic, which implies 
linear well-posedness. 

We inferred that moment systems can alternatively be conceived of as Galerkin 
approximations of a renormalized Boltzmann equation. We considered moment systems
based on a renormalization map composed of Tsallis' $q$-exponential. This renormalization map is concomitant with a $\varphi$\nobreakdash-divergence corresponding
to the anti-derivative of the inverse $q$\nobreakdash-exponential, which yields a natural approximation to relative entropy. The evaluation of moments of $q$\nobreakdash-exponential, elementary in numerical methods for the corresponding moment system, is tractable,
as opposed to the evaluation of moments of exponentials of arbitrary-order polynomials, 
connected with maximum-entropy closure. 

Numerical results have been presented for the one-dimensional spatially homogeneous Boltzmann-BGK equation. The nonlinear projection problem associated with the moment-constrained
$\varphi$\nobreakdash-divergence minimization problems was solved by means of Newton's
method. We observed that the condition number of the Jacobian matrices in the tangent
problems generally deteriorates as the number of moments increases. Nevertheless, in
all considered cases approximations up to at least 14 moments could be computed. We
observed that the $q$\nobreakdash-exponential approximation converges exponentially 
in the $L^1(\IR)$-norm with increasing number of moments. Moreover, we demonstrated
that functionals of the approximate distribution display super convergence, in accordance
with the Babu\v{s}ka--Miller theorem for Galerkin approximations.

\appendix

\section{Generalized BGK collision operator}
\label{Appendix:GBGK}
In \cite{Levermore1996}, Levermore introduced a class of multiscale 
generalizations of the BGK collision operator based on a finite sequence of 
increasingly constrained entropic projections of the form~(\ref{eq:EntMin}). 
In particular, given an admissible space of polynomials~$\IM$, consider a 
sequence of nested subspaces $\{\mathbb{M}_k\}_{k=1}^K$ with $\IM_0=\IE$ and strictly contained in $\mathbb{M}$, i.e.
$\IE=\mathbb{M}_1\subset \mathbb{M}_2\subset ... \subset \mathbb{M}_K\subset
\mathbb{M}$.
For each~$k$ and $f\in\IF$, let $f\mapsto\mathcal{F}^k(f)=:\mathcal{F}^k$ 
be the $\IM_k$\nobreakdash-moments constrained entropic projection of~$f$,
\begin{equation}
\label{eq:EntProj}
\mathcal{F}^k(f):=
\argmin_{g\in\mathbb{F}}
\big\{\langle \eta(g)\rangle:\langle m g \rangle=\langle m f \rangle,\,\forall{}m\in\IM_k\big\},
\end{equation}
with $\eta(g)=g\log{}g-g$,
under the assumption that~\EQ{EntProj} admits a solution for each~$k$. Based on
the sequence of projections $\{\mathcal{F}^k\}_{k=1}^K$, one can define 
a multiscale relaxation operator:
\begin{equation}
\label{eq:GenBGK1}
\mathcal{C}(f) =
-\theta_K(f-\mathcal{F}^K)-\sum_{k=1}^{K-1}\theta_k(\mathcal{F}^{k+1}-\mathcal{F}^{
k})
\end{equation}
with $\{\theta_k\}_{k=1}^{K}$ an increasing sequence of positive relaxation
rates depending on~$f$. The relaxation rate $\theta_k$  with $k\in\{1,2,\ldots,K-1\}$
constitutes the rate at which 
$\mathcal{F}^{k+1}$ decays to $\mathcal{F}^{k}$, while $\theta_K$ is the rate at
which $f$ decays to $\mathcal{F}^K$. In~\cite{Levermore1996} it is shown that
the Prandtl number can be controlled via the relaxation rates.

The above construction of the generalized BGK operator can be extended to 
$\varphi$\nobreakdash-divergences. To this end, consider an 
arbitrary $\varphi$\nobreakdash-divergence and 
let $f\mapsto\mathcal{F}^k(f)=:\mathcal{F}^k$ denote
the corresponding divergence-minimization projection according to~\EQ{EntProj},
i.e. $\mathcal{F}^k$ is defined by~\EQ{EntProj} 
with $\eta(\cdot)=\MM\varphi((\cdot)/\MM)$. Based on the projections $\mathcal{F}^k$, an extended BGK operator can be defined analogous to~\EQ{GenBGK1}. To establish that
$\eta'(\cdot)=\varphi'((\cdot/\MM)$ corresponds to an entropy density for 
the generalized BGK operator, we first note that the (strong) convexity of~$\eta$ implies:
\begin{equation}
\label{eq:ConvexHH}
(\HH'(s)-\HH'(t))(s-t)\big\rangle \geq0
\end{equation}
for all $s,t$ in the domain of $\eta$ and equality in~\EQ{ConvexHH} holds if and only if $s=t$. Rearranging the sum in~\EQ{GenBGK1} yields:
\begin{equation}
\label{eq:GenBGK2}
\CC(f)=
-\theta_1\big(f-\mathcal{F}^1\big)
-\sum_{k=1}^{K-1}\big(\theta_{k+1}-\theta_{k}\big)\big(f-\mathcal{F}^{k+1}\big)
\end{equation}
From the minimization problem~\EQ{EntProj} we infer 
that for all $k$ it holds that~$\eta'(\mathcal{F}^k)\in\IM_k$ and $\langle{}m(f-\mathcal{F}^k)\rangle=0$ for all $m\in{}\IM_k$. Hence, $\langle\eta'(\mathcal{F}^k)f\rangle-\langle\eta'(\mathcal{F}^k)\mathcal{F}^k\rangle=0$ yields a partition of zero 
for all~$k$. From~\EQ{GenBGK2} and the aforementioned partition of zero, we obtain
\begin{equation}
\label{eq:HHCf}
\begin{aligned}
\big\langle\eta'(f)\CC(f)\big\rangle
&=
-\theta_1\big\langle\eta'(f)\big(f-\mathcal{F}^1\big)\big\rangle
-\sum_{k=1}^{K-1}\big(\theta_{k+1}-\theta_{k}\big)\big\langle\eta'(f)\big(f-\mathcal{F}^{k+1}\big)\big\rangle
\\
&=
-\theta_1\big\langle\big(\eta'(f)-\eta'(\mathcal{F}^1\big)\big(f-\mathcal{F}^1\big)\big\rangle
-\sum_{k=1}^{K-1}\big(\theta_{k+1}-\theta_{k}\big)\big\langle\big(\eta'(f)-\eta'(\mathcal{F}^{k+1})\big)\big(f-\mathcal{F}^{k+1}\big)\big\rangle
\end{aligned}
\end{equation}
From $\theta_1>0$ and $\theta_{k+1}>\theta_k$ ($k=1,\ldots,K-1$), and the
convexity of $\eta(\cdot)$ according to~\EQ{ConvexHH} we conclude that 
$\eta$ and $\mathcal{C}$ satisfy the dissipation relation~\EQ{Dissipation},
i.e. $\langle\eta'(f)\mathcal{C}(f)\rangle\leq{}0$ for all admissible~$f$.
To verify the second prerequisite relation between $\eta$ and~$\mathcal{C}$, viz., the equivalence of the statements in \EQ{Equilibrium}, we first observe that the
implication \EQ{Equilibrium}${}_{(i)}$ $\Rightarrow$ \EQ{Equilibrium}${}_{(ii)}$ 
is trivial. To validate the reverse implication in~\EQ{Equilibrium}, 
we note that \EQ{Equilibrium}${}_{(ii)}$ in combination with 
the convexity of~$\eta$ according to~\EQ{ConvexHH} 
and the ultimate expression in~\EQ{HHCf} implies that 
$\smash[t]{(\eta'(f)-\eta'(\mathcal{F}^k))(f-\mathcal{F}^k)}$ vanishes almost everywhere
for all $k=1,\ldots,K$.
This, in turn, implies that $f=\mathcal{F}^1=\cdots=\mathcal{F}^K$.  
Condition~\EQ{Equilibrium}${}_{(i)}$ then follows directly from~\EQ{GenBGK2}.
To verify the implication 
\EQ{Equilibrium}${}_{(ii)}$~$\Rightarrow$~\EQ{Equilibrium}${}_{(iii)}$, 
we note that $\smash[t]{\mathcal{F}^k}$ according to~\EQ{EntProj} satisfies
$\smash[t]{\eta'(\mathcal{F}^k)\in\IM_k}$
for all~$k$. Recalling that~\EQ{Equilibrium}${}_{(ii)}$ implies 
$f=\smash[t]{\mathcal{F}^1}$, we infer $\eta'(f)\in\IM_1=\IE$ in accordance 
with~\EQ{Equilibrium}${}_{(iii)}$. Finally, the reverse implication 
\EQ{Equilibrium}${}_{(iii)}$~$\Rightarrow$~\EQ{Equilibrium}${}_{(ii)}$
follows immediately from~\EQ{GenBGK2} and the moment constraints in~\EQ{EntProj}.

\bibliographystyle{plain}
\bibliography{BibFile}   

\end{document}